\documentclass[aps,prb,twocolumn,showpacs,showkeys,footinbib,superscriptaddress]{revtex4-2}

\usepackage{amsmath}
\usepackage{amssymb}
\usepackage{graphicx}
\usepackage{bm}
\usepackage{color}
\usepackage{relsize}
\usepackage{braket}

\usepackage{hyperref}
\usepackage[all]{hypcap}

\begin{document}
\title{Energy shifts and broadening of excitonic resonances in electrostatically-doped semiconductors}   
\date{\today}

\author{Hanan~Dery}
\email[]{hanan.dery@rochester.edu}
\affiliation{Department of Electrical and Computer Engineering, University of Rochester, Rochester, New York 14627, USA}
\affiliation{Department of Physics and Astronomy, University of Rochester, Rochester, New York 14627, USA}
\author{Cedric Robert}
\affiliation{Universit\'{e} de Toulouse, INSA-CNRS-UPS, LPCNO, 135 Av. Rangueil, 31077 Toulouse, France}
\author{Scott~A.~Crooker}
\affiliation{National High Magnetic Field Laboratory, Los Alamos, New Mexico 87545, USA}
\author{Xavier Marie}
\affiliation{Universit\'{e} de Toulouse, INSA-CNRS-UPS, LPCNO, 135 Av. Rangueil, 31077 Toulouse, France}
\author{Dinh Van Tuan}
\affiliation{Department of Electrical and Computer Engineering, University of Rochester, Rochester, New York 14627, USA}

\begin{abstract} 
Tuning the density of resident electrons or holes in semiconductors through electrostatic doping provides crucial insight into the composition of the many excitonic complexes that are routinely observed as absorption or photoluminescence resonances in optical studies. Moreover, we can change the way these resonances shift and broaden in energy by controlling the quantum numbers (e.g., spin and valley) of the resident carriers with applied magnetic fields and doping levels, and by selecting the quantum numbers of the photoexcited or recombining electron-hole (e-h) pair through optical polarization. Here, we discuss the roles of \textit{distinguishability} and \textit{optimality} of excitonic complexes, showing them to be key ingredients that determine the energy shifts and broadening of optical resonances in charge-tunable semiconductors. A  \textit{distinguishable} e-h pair means that the electron and hole undergoing photoexcitation or recombination have quantum numbers that are not shared by any of the resident carriers. An  \textit{optimal} excitonic complex refers to a complex whose particles come with all available quantum numbers of the resident carriers. Based on the carrier density, magnetic field, and light polarization, all optical resonances may be classified as either distinct or indistinct depending on the distinguishability of the e-h pair, and the underlying excitonic complex can be classified as either optimal or suboptimal.  The universality of these classifications, inherited from the fundamental Pauli exclusion principle, allows us to understand how optical resonances shift in energy and whether they should broaden as doping is increased. This understanding is supported by conclusive evidence that the broadening and decay of optical resonances cannot be simply attributed to enhanced screening when resident carriers are added to a semiconductor. Finally, applying the classification scheme in either monolayer or moir\'{e} heterobilayer semiconductor systems, we relate the energy shift and amplitude of the neutral exciton resonance to the compressibility of the resident carrier gas.
\end{abstract}


\maketitle

\section{Introduction}

Optical absorption and emission processes in electrostatically-doped semiconductors tell us a great deal about the pervading role of the Pauli exclusion principle. {\color{black}One ramification of the Pauli principle is that under most conditions, excitonic complexes in semiconductors are composed of particles with different valley and spin quantum numbers \cite{Jones_NatPhys16,Courtade_PRB17,Chen_NatComm18,Ye_NatComm18,Li_NatComm18,Barbone_NatComm18,Liu_PRL20,Li_NanoLett22,Mostaani_PRB23}.} In addition, the Pauli principle restricts the permissible types of recombination channels of dark excitons and trions \cite{He_NatComm20,Liu_PRL20b,Robert_PRL21,Li_PRB22}, the way hot excitons relax in energy  \cite{Honold_PRB89,Eccleston_PRB91,Kochereshko_PRB98,Ramon_PRB03,Shahnazaryan_PRB17,Yang_PRB22}, and how impurities assist recombination processes \cite{Ren_PRB23}. However, when it comes to the relation between electrostatic doping and the broadening of excitonic optical resonances, or between electrostatic doping and the energy shifts of these resonances  \cite{Wang_NanoLett17,Smolenski_PRL19,Wang_PRX20,Liu_NatComm21,VanTuan_PRL22,Choi_PRB24,Tang_Nature20,Wang_NatMater23}, the Pauli exclusion principle is often given a lesser role and other mechanisms are invoked to explain such physics. To date, no proposed mechanism can satisfactorily or self-consistently explain the broadening and energy shifts of all optical resonances  in charge-tunable semiconductors, including theoretical frameworks based on exciton-polarons \cite{Sidler_NatPhys17,Efimkin_PRB17,Efimkin_PRB18,Fey_PRB20,Huang_PRX23}, correlated trions (i.e. ``tetrons'') \cite{Bronold_PRB00,Suris_PSS01,Koudinov_PRL14,Chang_PRB18,Rana_PRB20,VanTuan_PRB22}, intervalley plasmons \cite{Dery_PRB16,VanTuan_PRX17,VanTuan_PRB19,Dery_PRB16}, or arguments that rely on band filling and screening \cite{Steinhoff_NatComm17,Scharf_JPCM19, Strinati_PRL82,Mhenni_ACSNano25}. 

To explain all experimental observations self-consistently and to inform on the essential role played by the Pauli exclusion principle, we propose in this work that all excitonic optical resonances should be classified as being either \textit{distinct} or \textit{indistinct}, and that their underlying excitonic complexes should be classified as being either \textit{optimal} or \textit{suboptimal}. {\color{black} We first provide formal definitions for the distinctiveness of resonances and the optimality of excitonic complexes. In the next sections and with the aid of the examples shown in Figs.~\ref{fig:MoSe2}-\ref{fig:hb}, we will show how to apply these definitions and how}  this classification scheme provides a useful framework to understand whether and how various excitonic resonances shift and broaden in energy as the charge density is varied.
 
 \vspace{2mm} 
\textit{Distinct versus indistinct optical resonances}:  
When a semiconductor is doped with resident electrons in the conduction band (CB) or resident holes in the valence band (VB), there is an important difference between optical transitions that involve distinguishable versus indistinguishable e-h pairs. A distinguishable pair means that the electron and hole have unique quantum numbers that are not shared by \textit{any} of the resident carriers. For example, when the electron and hole are photoexcited into unoccupied valleys. Distinct (indistinct) optical resonances arise from optical transitions of distinguishable (indistinguishable) e-h pairs. To streamline the discussion in the rest of this work, we will also refer to resident carriers as indistinguishable if they share quantum numbers with the electron or hole in the pair, and as distinguishable otherwise.  

\vspace{4mm} 
\textit{Optimal versus suboptimal excitonic complexes}: 
We say that an excitonic complex is optimal if there are no resident carriers in the system with which the complex could bind to form a different lower-energy complex with more particles.  {\color{black} Other than in very special circumstances (Appendix~\ref{sec:caveats})}, particles with similar quantum numbers cannot be part of the same bound complex {\color{black} \cite{Mostaani_PRB17,Tiene_PRB22,VanTuan_PRB25}}. As such, the electrons (or holes) of an optimal excitonic complex are mutually distinguishable and they include all possible valley-spin species of resident electrons (or holes) at the edge of the CB (VB). This definition implies that the ground-state excitonic complex is always optimal. However, we will show that other complexes can be optimal as well.

\vspace{2mm}

We will employ these definitions and present an overarching framework by which one can directly analyze the measured behavior of all optical resonances in charge-tunable semiconductors. Focusing on transition-metal dichalcogenide (TMD) monolayers, their moir\'{e} heterobilayers, {\color{black} and conventional III-V and II-VI semiconductor quantum wells (QWs)} by using the experimental findings of Refs.~\cite{Liu_PRL20,Li_NanoLett22,Robert_PRL21,Wang_NanoLett17,Smolenski_PRL19,Wang_PRX20,Liu_NatComm21,VanTuan_PRL22,Choi_PRB24,Tang_Nature20,Wang_NatMater23,Kheng_PRL93,Astakhov_PRB00,Shields_PRB95}, we identify five important understandings and universal observations. 

\begin{enumerate}

\item \textit{Distinct} resonances of \textit{optimal} excitonic complexes neither broaden nor decay when resident carriers are added to the semiconductor. Examples include certain types of trions in {\color{black}valley- or spin-polarized semiconductors}, 6-particle hexcitons, or intralayer excitons in moir\'{e} heterobilayers. These resonances only show a steady energy redshift that is likely caused by a small  difference between renormalizations of the band gap and binding energies. There is no broadening or decay in this case because the absorption or recombination processes do not conflict with Pauli exclusion and therefore there is no need to shake-up the Fermi sea (i.e., to rearrange the resident carriers).

\item \textit{Indistinct} resonances of \textit{optimal} excitonic complexes do broaden and slowly shift in energy due to shakeup processes when the charge density increases. Examples include certain trions in zero magnetic field. 

\item While Coulomb screening by resident carriers gives rise to the small difference between renormalizations of the band gap and binding energies, we present conclusive evidence that this screening does not lead to the decay and broadening of any resonances in the optical spectra.

\item With increasing charge density, the energy blueshift and broadening of the neutral-exciton absorption resonance are especially strong in the presence of distinguishable resident carriers. {\color{black} The energy blueshift stems from transferring the spectral weight of the suboptimal exciton to the optimal trion resonance. The resulting energy blueshift is stronger than the Moss–Burstein effect \cite{Moss_PPSB54,Burstein_PR54}, which is caused by Pauli blocking (band filling) when indistinguishable resident carriers occupy low energy states in the photoexcited valley.} 

\item The energy shift and amplitude of the neutral exciton resonance are related to the compressibility of the resident carrier gas. We explain the unique behavior of this resonance when the resident carriers form an incompressible many-body state, including integer filling of Landau levels in a monolayer subjected to a strong magnetic field or fractional filling of a moir\'{e} lattice at certain doping levels.

\end{enumerate}

Before embarking, it is important to state that our analysis is valid in cases where the neutral exciton binding energy is much larger than the Fermi energy or cyclotron energy when applying a magnetic field, $\varepsilon_X \gg E_F,\,\hbar\omega_c$, meaning that the effective Bohr radius of the exciton is smaller than the cyclotron radius or the average distance between resident carriers. 

\subsection*{Organization of the paper} 

Section~\ref{sec:do} outlines the classification of optical resonances and excitonic complexes in MoSe$_2$ and WSe$_2$ monolayers based on their distinctiveness and optimality. Section~\ref{sec:obs} discusses various experimental results and then focuses on the universal energy redshift of distinct resonances of optimal excitonic complexes when the charge density increases in {\color{black}TMD monolayers, moir\'{e} heterobilayers, and conventional semiconductor QWs}. We then show that screening by resident carriers is not leading to the decay of excitonic complexes.  Section~\ref{sec:shakeup} deals with shakeup processes that correspond to the broadening and energy shifts of indistinct resonances of optimal excitonic complexes. {\color{black} Section~\ref{sec:blueX0} analyzes the energy blueshift of photoexcited neutral excitons under various absorption conditions}. Section~\ref{sec:point2} deals with the relation between the compressibility of the resident carrier gas and the behavior of neutral exciton resonances in monolayers and moir\'{e} heterobilayers.  A summary is given in Sec.~\ref{sec:outlook}. {\color{black} Appendices~\ref{sec:caveats}-\ref{sec:eh_exc} elaborate on key conclusions made in the main text and they provide technical details of calculated results.} 

 \section{classification of optical resonances and excitonic complexes}\label{sec:do}

\begin{figure} 
\centering
\includegraphics[width=8.3cm]{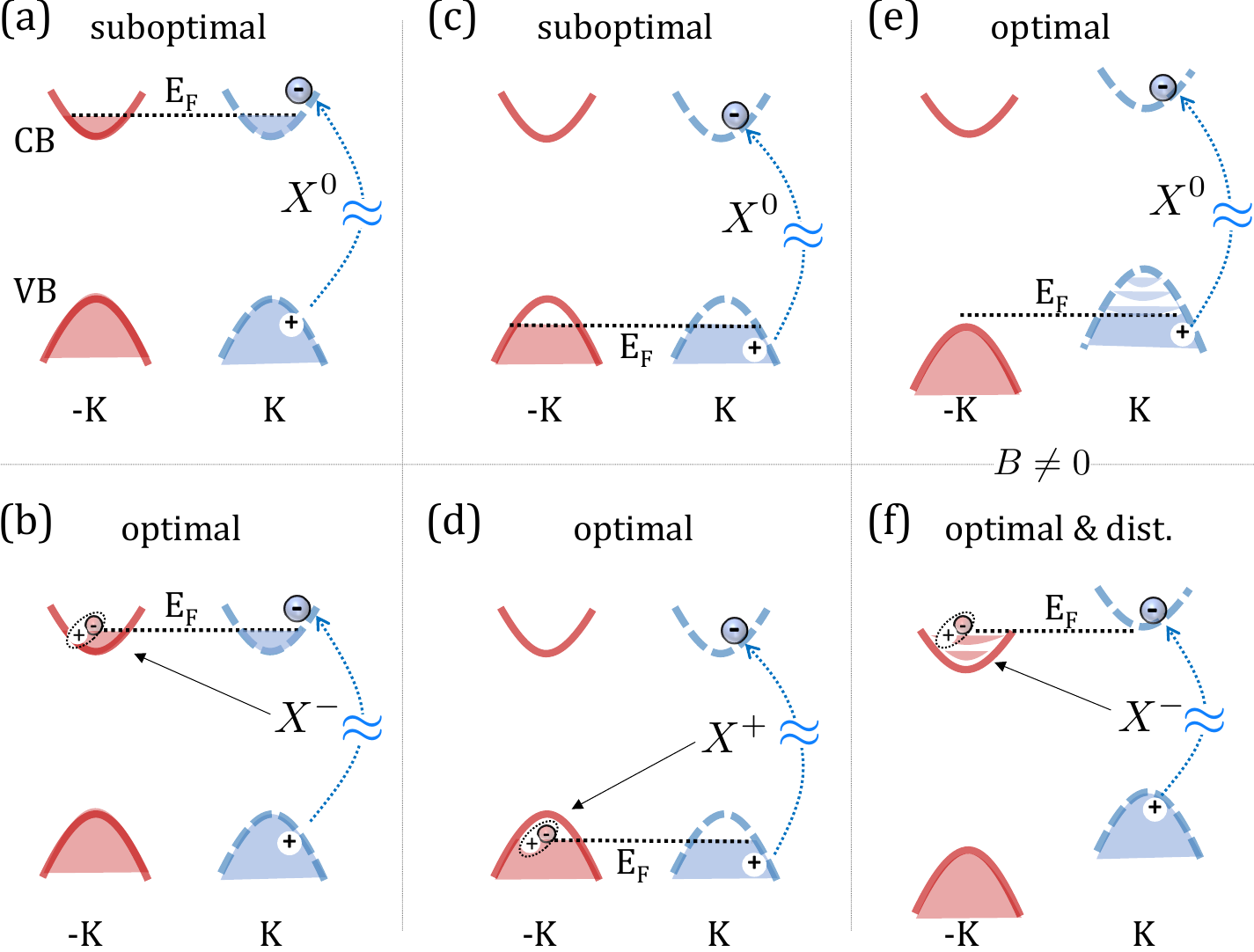}
\caption{Photoexcitation of the $K$ valley in an electrostatically-doped MoSe$_2$ monolayer. Valleys with solid (dashed) lines host spin-up (spin-down) electrons. The diagrams describe neutral excitons ($X^0$) and trions ($X^{\pm}$). For the latter, the photoexcited e-h pair binds to a charge particle from the $-K$ valley, shown as a particle-hole pair of the Fermi sea. The diagrams in (e)-(f) describe photoexcitation under a strong magnetic field $B$, such that Landau quantization of VB holes (CB electrons) happens solely at the $K$ ($-K$) valley. All optical transitions except for (f) are indistinct resonances, because the photoexcited electron or hole is injected into already-occupied bands.}\label{fig:MoSe2} 
\end{figure}

Figure~\ref{fig:MoSe2} shows examples of exciton and trion optical transitions in electrostatically-doped MoSe$_2$ monolayers, and how these  transitions and the underlying excitonic complexes are classified.  We choose to start with MoSe$_2$ for its simplicity, where optical transitions are between the upper VBs and lower CBs in both the $K$ and $-K$ valleys. The upper spin-orbit-split CBs (not shown) are assumed to be unpopulated and hence play no role.  The carrier populations in the upper VBs and lower CBs are controlled by doping level and by applied magnetic fields $B$. Without loss of generality, we discuss the resonances and exciton complexes as arising from optical transitions in the $K$ valley (i.e., with $\sigma_+$ circularly polarized light). 

Figures~\ref{fig:MoSe2}(a) and (b) show the photoexcitation in electron-rich conditions. Both exciton and trion resonances are said to be indistinct because the photoexcited electron is promoted to a CB that is already occupied by resident electrons having the same quantum numbers (spin and valley). Moreover, the neutral exciton in Fig.~\ref{fig:MoSe2}(a) is suboptimal because there are opposite-spin electrons in the $-K$ CB with which this exciton could bind to form a lower-energy complex with more particles (i.e., a trion).  Conversely, the negative trion shown in Fig.~\ref{fig:MoSe2}(b) is optimal because the photoexcited pair binds to an electron from the valley at $-K$, and there are no other kinds of resident electrons with which the pair can bind. The trion is depicted as the binding between the photoexcited e-h pair and particle-hole pair from the Fermi sea in the $-K$ valley, and the entire complex is known as a tetron, wherein the trion moves together and is correlated with the CB hole in the Fermi sea around the trion \cite{Bronold_PRB00,Suris_PSS01,Koudinov_PRL14}. Figures~\ref{fig:MoSe2}(c) and (d) show the corresponding transitions in a hole-rich MoSe$_2$ monolayer. 

Applying a magnetic field $B$ can change the classifications if $B$ is strong enough to fully valley-polarize the Fermi sea of resident carriers. In the hole-rich case, shown in Fig.~\ref{fig:MoSe2}(e), the absence of resident VB holes in the $-K$ valley (with which the neutral exciton could otherwise bind) renders $X^0$ an optimal complex, although its resonance remains indistinct.  In the electron-rich case, shown in Fig.~\ref{fig:MoSe2}(f), the absence of resident CB electrons at the $K$ valley of the photoexcited electron causes the resonance of this optimal trion to become distinct.  

\begin{figure} 
\centering
\includegraphics[width=8.3cm]{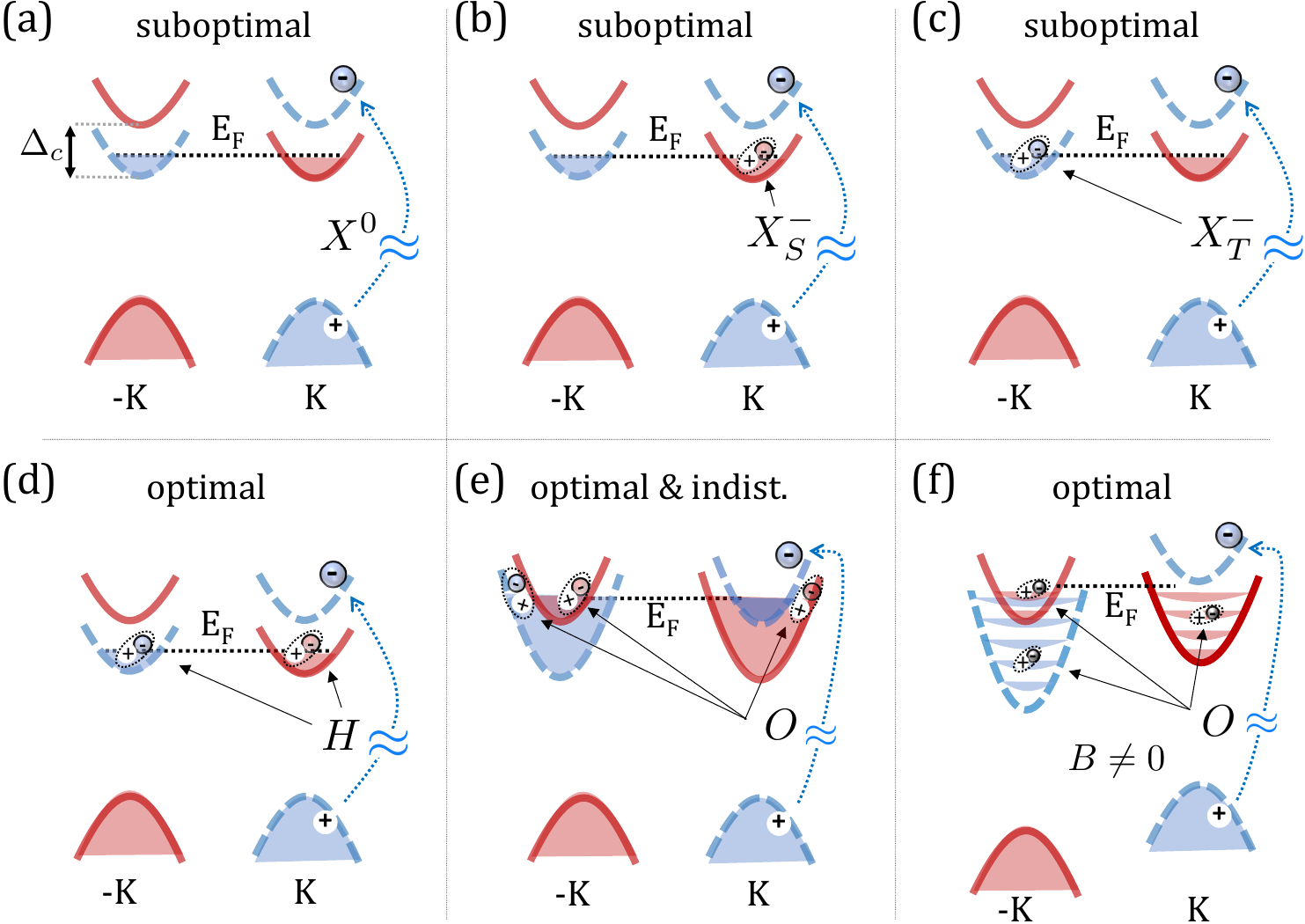}
\caption{Photoexcitation of the $K$ valley in an electron-rich WSe$_2$ monolayer, where optically-allowed transitions couple to the upper CBs. Valleys with solid (dashed) lines host spin-up (spin-down) electrons. (a)-(d) Photoexcitation of charge-neutral excitons ($X^0$), negative trions ($X^-_{S,T}$), and hexcitons ($H$) when the upper CBs are not populated. (e)-(f)  Photoexcitation of oxcitons ($O$) when the upper CBs are populated, without and with a magnetic field. All optical transitions except for (e) are distinct resonances.}\label{fig:WSe2} 
\vspace{-2mm}
\end{figure}

The excitonic complexes in WSe$_2$ (or WS$_2$) monolayers are similar in character to those of MoSe$_2$ for the case of hole doping, but are qualitatively different for the case of electron doping.  This is because the sign of the CB spin-orbit splitting is opposite in W-based TMD monolayers in comparison to MoSe$_2$. As shown in Fig.~\ref{fig:WSe2}, photoexcited electrons are promoted to the upper CBs, rendering the photoexcited e-h pair distinguishable in electron-rich conditions as long as the upper CBs are not populated by resident electrons.  The photoexcitation involves distinct resonances for the neutral exciton in Fig.~\ref{fig:WSe2}(a), singlet trion in (b), triplet trion in (c), and the optimal hexciton complex in (d). The hexciton comprises six particles: the photoexcited e-h pair, and two particle-hole pairs from the two lower CBs at $\pm K$ \cite{Choi_PRB24,VanTuan_PRL22,VanTuan_PRB22} (Appendix~\ref{sec:hex}).  When the upper CB valleys start to fill at high doping levels, the photoexcited e-h pair can form an eight-particle oxciton by binding to an additional particle-hole pair \cite{VanTuan_PRL22}, as shown in Fig.~\ref{fig:WSe2}(e). The oxciton becomes the optimal complex but it has an indistinct resonance at $B$=0. A strong magnetic field can turn this resonance distinct again by emptying the upper CB at $K$, as shown in Fig.~\ref{fig:WSe2}(f). 

\subsection*{Type-B exciton complexes at higher energies}\label{sec:typeB}

\begin{figure}[t]
\centering
\includegraphics[width=8cm]{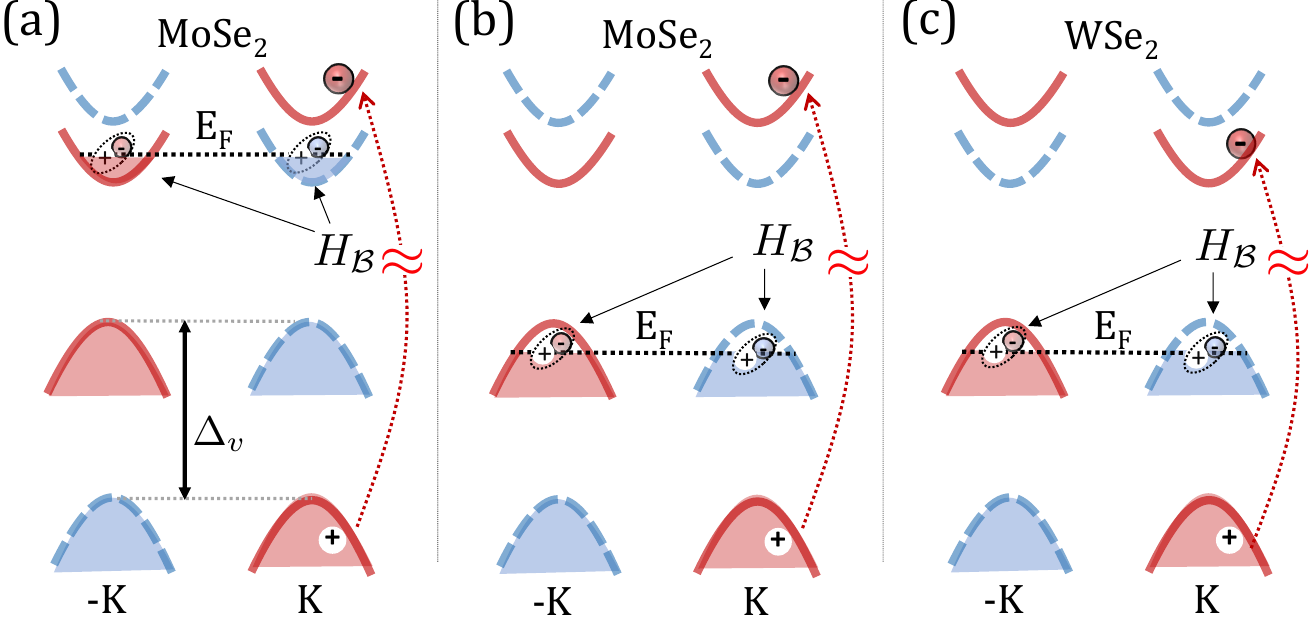}
\caption{Photoexcitation of type-B hexcitons in (a) electron-doped MoSe$_2$, (b) hole-doped MoSe$_2$, and (c) hole-doped WSe$_2$ (or WS$_2$) monolayers. All of these six-particle hexcitons are optimal complexes (because the photoexcited e-h pair couples to all available species of resident carriers), and their optical resonances are distinct (because the e-h pair is photoexcited into empty bands).} \label{fig:hb} 
\end{figure}

Thus far, we have discussed optical transitions originating from the uppermost VBs at $\pm K$, which are the well-known ``type-A'' excitonic resonances that appear on the low-energy side of the optical spectrum. We emphasize that the type-A resonance with lowest energy (the ground state) always belongs to an optimal excitonic complex. However, excitonic resonances also appear at much higher energies when the photoexcited hole originates from the lower of the spin-orbit-split VBs; these are the ``type-B'' exciton complexes.  Their energies exceed those of type-A resonances by approximately the spin-orbit splitting of the VBs: $\Delta_v$$\,\sim$180~meV in Mo-based TMDs and $\Delta_v$$\,\sim$450~meV in W-based TMDs \cite{Kormanyos_2DMater15}. Figure~\ref{fig:hb} shows examples of distinct optical resonances of type-B optimal hexcitons. These complexes emerge in hole-rich TMD monolayers (Mo- and W-based) or in electron-rich MoSe$_2$ monolayers when the upper CBs are unpopulated \cite{Choi_PRB24}. 

\begin{figure*}[tbh] 
\centering
\includegraphics[width=18cm]{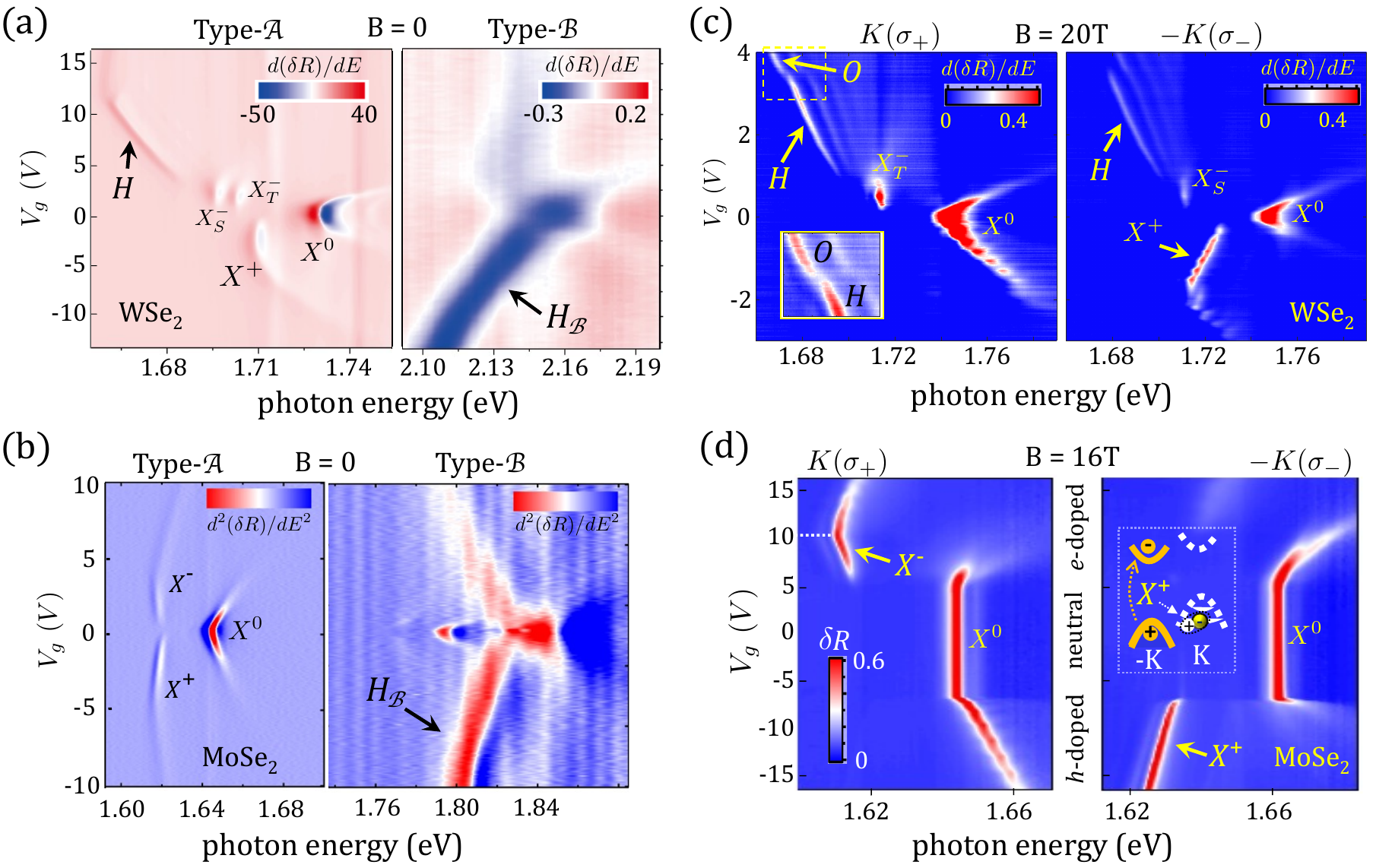}
\caption{(a) Colormaps of the type-A and type-B optical resonances from gate-dependent reflectance spectra in WSe$_2$ at zero magnetic field. Taken from Ref.~\cite{Wang_NanoLett17}. (b) The same, but for MoSe$_2$. Taken from Ref.~\cite{Liu_NatComm21}. (c) Helicity-resolved magneto-optical reflectance spectra in WSe$_2$ when the applied magnetic field is 20~T. Taken from Ref.~\cite{Wang_PRX20}. Inset: magnified view of the hexciton and oxciton resonances in the dashed box at the top left corner of the colormap. (d) The same as in (c) but for MoSe$_2$ at 16~T. Taken from Ref.~\cite{Smolenski_PRL19}. Inset: band diagram of $X^+$. In all colormaps, distinct resonances of optimal excitonic complexes are indicated by arrows.  In (a)-(c), $V_g \approx 0$ corresponds to charge neutrality, $V_g > 0$ to electron doping, and $V_g < 0$ to hole doping. The corresponding regimes in (d) are  $-7$V$\,\lesssim\,$$V_g \lesssim 6$V, $V_g \gtrsim 6$V and $V_g \lesssim -7$V.}\label{fig:dr} 
\end{figure*}

\section{Classification of excitonic complexes in experiment and their universal trends}\label{sec:obs}

Figure~\ref{fig:dr} shows colormaps of the low-temperature reflectance spectra in electrostatically-doped WSe$_2$ and MoSe$_2$ monolayers, taken from Refs.~\cite{Wang_NanoLett17,Liu_NatComm21,Wang_PRX20,Smolenski_PRL19}.  Spectral windows that correspond to type-A and type-B optical transitions at zero magnetic field are shown in Fig.~\ref{fig:dr}(a) for WSe$_2$ ~\cite{Wang_NanoLett17} and in Fig.~\ref{fig:dr}(b) for MoSe$_2$ \cite{Liu_NatComm21}. Helicity-resolved reflectance spectra of type-A optical transitions are shown in Fig.~\ref{fig:dr}(c) for WSe$_2$ at $B=20\,$T \cite{Wang_PRX20} and in Fig.~\ref{fig:dr}(d) for MoSe$_2$ at $B=16\,$T \cite{Smolenski_PRL19}. In all of these cases, \textit{distinct} resonances of \textit{optimal} excitonic complexes are indicated with arrows. These (and only these) resonances exhibit a pronounced and sustained energy redshift when the charge density increases. Below, we explain what can be learned from these measurements, and elaborate on universalities of the underlying physics. 

\vspace{-2mm}
\subsection{Universal energy redshift of distinct resonances of optimal excitonic complexes}\label{sec:redshift} 

We first focus on the energy redshift of distinct resonances that arise from optimal excitonic complexes. At zero magnetic field, these complexes are type-A hexcitons in electron-rich W-based monolayers and type-B hexcitons in any hole-rich TMD monolayer. Their resonances are marked by arrows in Figs.~\ref{fig:dr}(a) and (b), and their transition diagrams are shown in Figs.~\ref{fig:WSe2}(d), \ref{fig:hb}(b), and \ref{fig:hb}(c). Figure~\ref{fig:dr}(a) shows that the type-A hexciton resonance in WSe$_2$ remains sharp and exhibits a pronounced redshift with increasing electron density, which abruptly ceases only when the gate voltage reaches $\sim$11~V (electron density $\sim${\color{black}6}$\times$10$^{12}$~cm$^{-2}$) and electrons begin filling the upper CBs \cite{Wang_NanoLett17}. Beyond this point the photoexcited e-h pair becomes indistinguishable from some carriers in the Fermi sea, and the resonance therefore becomes indistinct. On the other hand, type-B hexcitons in hole-doped conditions remain optimal and their resonances remain distinct even at very large hole densities because $\Delta_v \gg E_F$, as depicted in Figs.~\ref{fig:hb}(b) and (c). These resonances, labeled by $H_{\mathcal{B}}$ in Figs.~\ref{fig:dr}(a) and (b), exhibit a robust redshift and do not broaden even up to the highest reported doping. We note that type-B resonances are typically broader than those of type-A, due to their very short lifetime \cite{footnote}.

Further important insight is obtained by applying magnetic fields $B$ large enough to fully valley-polarize the Fermi sea of resident carriers.  This forces certain optimal $X^\pm$ trion resonances, which are otherwise indistinct at $B$=0, to become distinct (because the e-h pair is now photoexcited into empty bands; see Fig.~\ref{fig:MoSe2}(f)). This completely changes how they shift and broaden with increasing doping. Figures~\ref{fig:dr}(c) and (d) show helicity-resolved reflectance spectra in WSe$_2$ and MoSe$_2$, respectively, at strong $B$ \cite{Wang_PRX20,Smolenski_PRL19}. The left and right panels show reflectance of $\sigma_+$ and $\sigma_-$ circularly polarized light, which probes the $K$ and $-K$ valleys,  respectively. To analyze the data we must consider the balance between Zeeman and Fermi energies. Once the monolayer becomes doped, a change of 1~V in gate voltage ($V_g$)  corresponds to a carrier density change of $\sim$1.5$\times$10$^{12}$~cm$^{-2}$ in the WSe$_2$ device \cite{Wang_PRX20} and $\sim$2.4$\times$10$^{11}$~cm$^{-2}$  in the MoSe$_2$ device \cite{Smolenski_PRL19}. In addition, the signs of the $g$-factors are opposite in the $\pm K$ valleys, where $|g_v| \approx 6$ for the VB, $|g_{c,2}| \approx 4$ for the CB which takes part in the type-A transition, and $|g_{c,1}| \approx 1$ for the CB which takes part in the type-B transition \cite{Robert_PRL21}. These $g$-factors mean that hole-rich monolayers can sustain complete valley polarization at larger carrier densities compared with electron-doped MoSe$_2$ monolayers ($g_v > g_{c,2}$), which in turn can sustain complete valley polarization at larger carrier densities compared with electron-doped W-based monolayers ($g_{c,2} > g_{c,1}$).

Starting with MoSe$_2$ in Fig.~\ref{fig:dr}(d), the $X^\pm$ resonances in $B$=16~T exhibit a pronounced redshift with increasing doping (in marked contrast to their negligible redshift at $B$=0).  However, this redshift persists \textit{only as long} as the optical transition involves a distinguishable e-h pair. The redshift of $X^-$ starts from its emergence when the monolayer becomes electron-doped at $V_g \sim$6~V and stops at $V_g \sim$10~V (indicated by the white dotted line near $X^-$ in Fig.~\ref{fig:dr}(d)). Beyond this point, the Fermi energy exceeds the Zeeman energy and the photoexcited valley starts to fill, and the $X^-$ resonance is no longer distinct, marking the onset of its broadening and energy blueshift. {\color{black}This behavior is consistent with other observations seen at different magnetic fields and electron densities \cite{Oreszczuk_2DMater13}.} Referring to the diagram of Fig.~\ref{fig:MoSe2}(f), the redshift ceases and the broadening is enhanced when the electron density is large enough to begin filling the CB valley at $K$.

For the case of $X^+$, whose energy diagram is shown by the inset of Fig.~\ref{fig:dr}(d), the redshift of its distinct resonance persists from its emergence when the monolayer becomes hole-doped at $V_g \sim -7$~V, and does not stop even at $V_g \sim -16$~V (hole density $\sim$2.2$\times$10$^{12}$~cm$^{-2}$). The larger value of $g_v$ \cite{Li_PRL20,Xuan_npj21} keeps the photoexcited valley free of resident holes across this entire range of gate voltages.

Similarly, Figure~\ref{fig:dr}(c) shows low-temperature differential reflectance spectra of a WSe$_2$ monolayer when $B=20$~T \cite{Wang_PRX20}. The $X^+$ trion is optimal and its resonance is distinct between 0 and $-1.7$~V (hole density up to $\sim$2.5$\times$10$^{12}$~cm$^{-2}$ in this device), during which its redshift is continuous and the resident holes are fully valley polarized. However, this redshift abruptly ceases when the resonance becomes indistinct at larger $V_g$ when holes begin populating Landau levels (LLs) in the $-K$ valley ($V_g \lesssim -1.7$~V), which also marks the onset of its enhanced broadening.  

Finally, we focus on the electron-doped side of Fig.~\ref{fig:dr}(c), where the gate voltage window of fully valley-polarized electrons is narrowest in W-based monolayers because of the smallness of $g_{c,1}$. The emergence of hexciton resonances  ($H$) and the concomitant decay of trions ($X^-_{S,T}$) takes place when LLs are filled in the lower CB valleys of both $\pm K$ \cite{VanTuan_PRL22}. Similar to the case at $B=0$ in Fig.~\ref{fig:dr}(a), the energy redshift of the type-A hexciton resonance persists only as long as the upper CB valleys are empty (i.e., as long as it remains a distinct resonance). The crossover from hexciton to oxciton occurs when the upper CB valley at $-K$ starts to fill, i.e. when $V_g \gtrsim 3.5$~V in this device (electron densities $\gtrsim${\color{black}6}$\times$10$^{12}$~cm$^{-2}$). The diagram of this distinct and optimal oxciton is shown in Fig.~\ref{fig:WSe2}(f). The observed $H\,\rightarrow\,O$ crossover is highlighted in the inset of Fig.~\ref{fig:dr}(c), which magnifies the reflectance signal with helicity $\sigma_+$ inside the dashed box at the top left corner (Appendix \ref{sec:hob}).

 \subsection{Suggested origin of the universal energy redshift}\label{sec:bgr}

These data show that distinct resonances of optimal excitonic complexes exhibit a universal energy redshift when the charge density increases, and further, that these resonances neither decay nor broaden. The lack of decay and broadening can be reasoned by noting that a distinguishable e-h pair can bind to resident carriers with all available quantum numbers \textit{without} having to perturb the distribution of resident carriers that are not part of the optimal complex (Appendix \ref{sec:hex}). In Sec.~\ref{sec:shakeup}, we will discuss why and how shakeup processes broaden the optical resonances in cases that the e-h pair is indistinguishable. 

The universal energy redshift is common to distinct resonances of various optimal complexes. This commonality can be attributed to the small but non-zero difference that exists between band-gap renormalization (BGR) and reduction of the binding energy due to screening by resident carriers (namely, $\Delta E = \Delta E_g - \Delta E_b$). Increasing the charge density from $n_1$ to $n_2=n_1+\delta n$ is equivalent to increasing the screening through the effective dielectric constant from $\epsilon_1$ to $\epsilon_2 = \epsilon_1 + \delta \epsilon$. The BGR of a semiconductor is attributed to the {\color{black}change in energy needed to excite an electron across the band gap when the Coulomb potential at the immediate vicinity of the electron is changed \cite{Marauhn_PRB23}}. Considering a small change, such that $\delta \epsilon$ is small and positive, we get \cite{Haug_SchmittRink_PQE84,Steinhoff_NanoLett14,Raja_NatComm17,Raja_NatNano19,Marauhn_PRB23}  
 \begin{eqnarray}
\Delta E_g &=&   E_{g,2} - E_{g,1} =  \lim_{r \rightarrow 0} \left[V(\epsilon_2, {\bf r}) - V(\epsilon_1, {\bf r})\right] \nonumber \\ &=&  \delta \epsilon \,\,  V' (\epsilon_1, r = 0 ), \label{eq:eg}  
\end{eqnarray}
where $V'(\epsilon_1, r)  = (\partial V / \partial \epsilon)\big|_{\epsilon=\epsilon_1}$. The exciton state is calculated with the same Coulomb potential. Using perturbation theory to quantify the change in exciton binding energy, we get
\begin{equation}
\Delta E_b = \delta \epsilon \langle\,\,  \psi_1 | V' (\epsilon_1, r ) | \psi_1 \rangle   ,
\end{equation}
where $|\psi_1\rangle$ is the exciton wave function corresponding to $V(\epsilon_1, r)$. The total shift of the exciton resonance energy becomes
\begin{eqnarray}
\Delta E_{X_0} &=& \Delta E_g - \Delta E_b \nonumber \\ &=&   \delta \epsilon \left\langle\,\,  \psi_1 \left| { \left[ V' (\epsilon_1, r = 0 ) -  V' (\epsilon_1, r ) \right]  } \right| \psi_1 \right\rangle.  
\end{eqnarray}
Assuming small size exciton, the term in square brackets is merely the derivative of $r$, and we get 
\begin{equation} 
\Delta E_{X_0} \simeq   -\delta \epsilon \left\langle\,\,  \psi_1 \left|   ({\partial^2 V}/{\partial \epsilon  \partial r}) \, r  \right| \psi_1 \right\rangle  < 0 . \label{eq:red}
\end{equation}
The energy change is negative because the potential $V(\epsilon, r)$ is largely proportional to  $1/\epsilon r$, so that  $\partial^2 V/ \partial \epsilon  \partial r >0$. In other words, the exciton energy \textit{redshifts} under a small increase in the effective dielectric constant (caused by a small increase in charge density). The energy redshift is expected since the reduction in band-gap energy comes from weakening of the Coulomb potential at $r \rightarrow 0$. On the other hand, the electron-hole binding energy is affected by the change to the Coulomb potential at finite distances and not only at $r \rightarrow 0$. Since the difference between Coulomb potentials with $\epsilon_1$ and $\epsilon_2$ is largest at $r=0$, the reduction in band-gap energy cannot be completely offset by weaker binding energy, leading to a small overall redshift of the resonance energy when the effective screening increases. Since the energy redshift is a result of increased effective screening, similar behavior is observed when monolayer semiconductors are sandwiched between materials with larger dielectric constants or when semi-metallic thin layers, such as graphite, are placed adjacent to the semiconductor \cite{Raja_NatComm17,Waldecker_PRL19}. 

For simplicity, the derivation above was made for a neutral exciton. Yet, the energy redshift of complexes with larger number of particles should be similar {\color{black}for two reasons. First, the energy shift that one observes in experiment is a measure of the change to the optical gap of the photoexcited (or recombined) electron-hole pair, whereas changes in the self-energy   of other particles in the complex are not probed in these optical experiments. For example, a trion inherits the self-energy of a resident carrier to which the e-h pair binds, and therefore, the absorbed photon energy cannot inform us on a self-energy that remains the same before and after the photoexcitation. More on this subtle point is discussed in Appendix~\ref{sec:optgap}. The second reason that the redshift is largely independent of the number of particles in the complex is that} most of the binding energy comes from the photoexcited pair, where adding particles contributes less and less energy. For example, if we consider a WSe$_2$ monolayer that is encapsulated in hexagonal boron nitride, then the gained energy by forming an exciton is about 170~meV compared with an unbound electron-hole pair \cite{Stier_PRL18}. Adding an electron to form a trion gains extra $\sim$30~meV \cite{Jones_NatPhys16}, adding the next electron to form a hexciton gains extra $\sim$10~meV, and the next electron to form an oxciton gains extra 3~meV \cite{Choi_PRB24,VanTuan_PRL22}.  This behavior is similar to that of energy levels of successive electron shells in atoms but with smaller energy scales.

\subsection{Optimality and distinguishability of neutral excitons in moir\'{e} heterobilayers} \label{sec:point2}

\begin{figure}
\includegraphics[width=8.5cm]{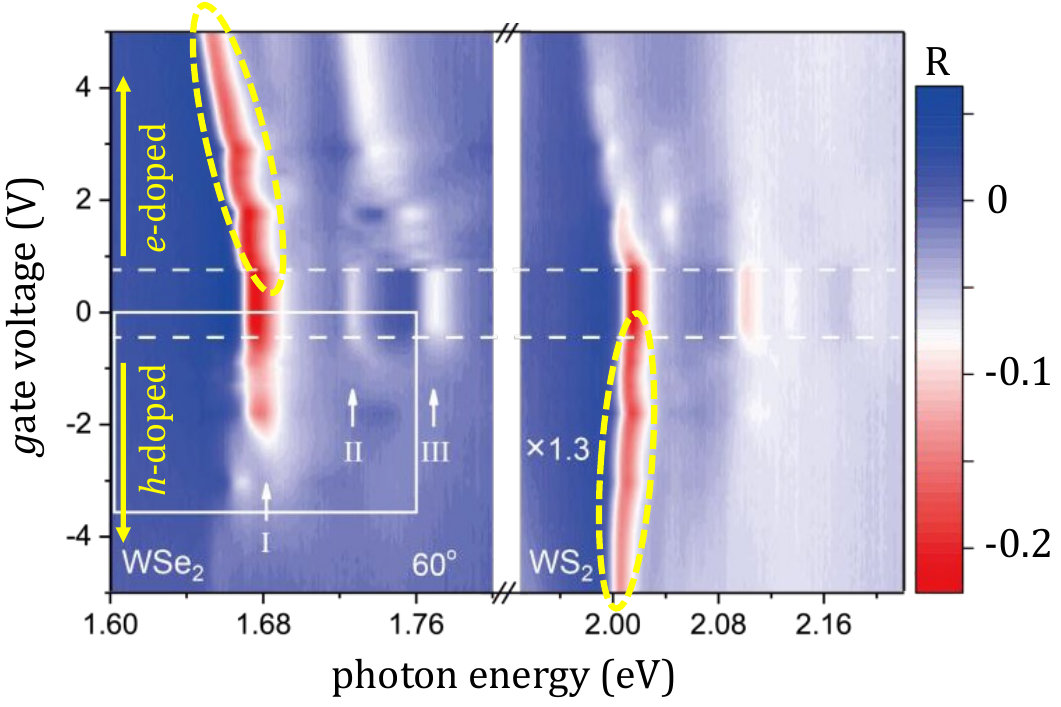}
\caption{Reflectance contrast spectrum of a WSe$_2$/WS$_2$ heterobilayer at low temperatures where the twist angle between the monolayers is 60$^\circ$. Taken from Ref.~\cite{Tang_Nature20}. The left and right panels show spectral regions of intralayer excitons in the WSe$_2$  and WS$_2$ monolayers, respectively.  The dashed-yellow ovals highlight regions in which the intralayer neutral excitons are optimal and their resonances distinct, supported by their sustained energy redshift and persistent amplitude at elevated charge densities.}\label{fig:moireR} 
\end{figure}

The concepts of optimality and distinguishability are universal and can also be applied to heterostructure systems. Unless we deal with the trivial intrinsic limit (i.e., no resident carriers), charge-neutral excitons in single monolayers cannot be both optimal and distinguishable. This restriction can be alleviated in heterobilayers where in addition to spin and valley, the monolayer in which the particle resides is a relevant quantum number. We illustrate this concept using the reflectance contrast spectrum of WSe$_2$/WS$_2$ moir\'{e} heterobilayers, shown in Fig.~\ref{fig:moireR}  \cite{Tang_Nature20}. The CB and VB alignments are such that electrons are added to the WS$_2$ monolayer when the gate voltage is positive, and holes are added to the WSe$_2$ monolayer when the gate voltage is negative. The low-energy spectral window, shown in the left part of Fig.~\ref{fig:moireR}, is governed by absorption processes in the WSe$_2$ monolayer.  The high-energy spectral window, shown in the right part, is governed by absorption processes in the WS$_2$ monolayer. 

To understand how a charge neutral exciton can be optimal and distinguishable in this heterobilayer system, we should exclude the possibility of creating trions with resident carriers. Since electrons reside in the WS$_2$ monolayer, an exciton in the WSe$_2$ monolayer is not affected by these resident electrons except for their screening-induced redshift effect from BGR and binding energy. The reason that a negative trion cannot form is the ineffective interaction between two electrons whose wavefunctions hardly overlap and a hole \cite{VanTuan_PRB25}. As such, the photoexcited e-h pair in the WSe$_2$ monolayer is an optimal exciton, and the lack of resident electrons in this monolayer renders the e-h pair  distinguishable. The intralayer neutral exciton in the WSe$_2$ monolayer at positive voltages is therefore an optimal complex with distinct resonance, whose hallmark is sustained energy redshift, as highlighted in the dashed yellow oval around 1.68~eV in Fig.~\ref{fig:moireR}. Here, adding resident electrons to the WS$_2$ monolayer screens the Coulomb potential in the WSe$_2$ monolayer, and the ensuing interplay between BGR and binding energy (Sec.~\ref{sec:bgr}) brings in the observed energy redshift. Equivalently, a positive trion cannot form when the neutral exciton resides in the  WS$_2$ monolayer and holes reside in the WSe$_2$ monolayer. The result is an optimal intralayer exciton in the WS$_2$ monolayer with distinct resonance, whose sustained energy redshift is highlighted in the dashed yellow oval around 2~eV in Fig.~\ref{fig:moireR}.

{\color{black}

\begin{figure*}[t!]
\centering
\includegraphics[width=17.5cm]{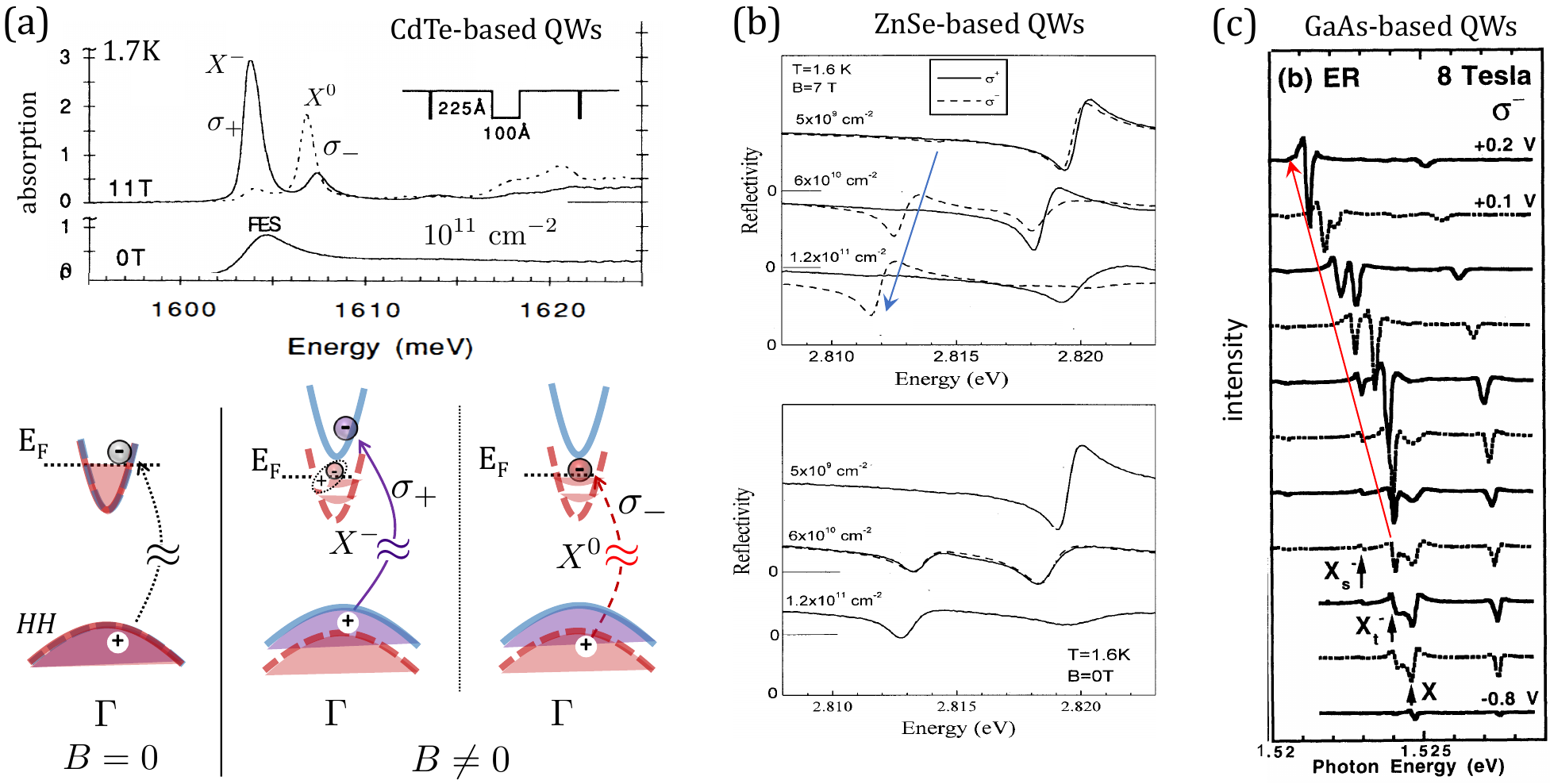}
\caption{{\color{black}Reflectance spectroscopy of semiconductor quantum wells (QWs). (a) Exciton and trion resonances in CdTe QWs at 1.7$\,$K. Taken from Ref.~\cite{Kheng_PRL93}. (b) The same but in ZnSe QWs at 1.6$\,$K. Taken from Ref.~\cite{Astakhov_PRB00}. (c) The same but in GaAs QWs at 2$\,$K. Taken from Ref.~\cite{Shields_PRB95}. The schemes in the bottom of (a) show zone-center optical transitions from the heavy-hole valence band, where their larger effective mass compared with electrons in the conduction band is illustrated by the valley parabolicity.}}\label{fig:QWs} 
\end{figure*}

\subsection{Optimality and distinguishability of excitons and trions in conventional semiconductor QWs}\label{sec:qws}

The proposed framework to classify excitonic complexes and their resonances is general and it extends beyond TMD monolayers and their moir\'{e} systems. Figure~\ref{fig:QWs} shows resonances of trions and excitons in absorption and reflectivity spectra of various semiconductor quantum wells (QWs) \cite{Kheng_PRL93,Astakhov_PRB00,Shields_PRB95}. Figure~\ref{fig:QWs}(a) shows the case of CdTe based-QWs at $1.7$~K when the electron density is 10$^{11}$~cm$^{-2}$, taken from Kheng \textit{et al}  \cite{Kheng_PRL93}. At $B=0$, the absorption spectrum includes a single broad resonance that the authors of Ref.~\cite{Kheng_PRL93} marked by FES (Fermi edge singularity). This singularity emerges because of the Coulomb enhancement in 2D systems \cite{SchmittRink_PRB86,Hawrylak_PRB91}, and its physics will be further discussed in Sec.~\ref{sec:shakeup}. Based on our proposed classification, we can argue that this resonance is an indistinct resonance of an optimal trion, which one would expect  from the optical transition in the left-bottom scheme of Fig.~\ref{fig:QWs}(a). Regardless of whether this is a FES resonance or an indistinct resonance of an optimal trion, the reflectance spectrum at $B=11$~T shows a different behavior under the same temperature and electron density. The resonances in this case are attributed to a negative trion with $\sigma_+$  polarization (solid line), and a charge-neutral exciton resonance with $\sigma_-$ polarization (dashed line). The magnetic field in this case is strong enough to fully polarize the electron gas, rendering the resonance of the optimal trion distinct, as shown by the bottom-middle scheme of Fig.~\ref{fig:QWs}(a). Compared with the resonance at $B=0$, the emergence of a narrower trion resonance  at lower energy (i.e., redshift) is consistent with the behavior of a distinct resonance of an optimal excitonic complex. Also in agreement with the expected behavior of optimal versus suboptimal excitons, the exciton resonance is observed at $B=11$~T because it becomes optimal, as one can realize from the bottom-right scheme of Fig.~\ref{fig:QWs}(a). A similar behavior is observed in TMD monolayers. For example, Figs.~\ref{fig:dr}(a) and (b) show that the resonance of the suboptimal exciton complex ($X^0$ at $B=0$) is rapidly suppressed when the charge density increases, whereas Figs.~\ref{fig:dr}(c) and (d) show that $X^0$ persists when the exciton becomes optimal at large fields (to be further discussed in Sec.~\ref{sec:blueX0}).

Figure~\ref{fig:QWs}(b) shows reflectivity spectra of ZnSe QWs at 1.6~K, taken from Ref.~\cite{Astakhov_PRB00}. The bottom and top panels show results at $B=0$ and 7~T, respectively, for three different charge densities. We notice two observations that are self-consistent with our framework. First, the exciton resonance at the larger density is amplified  when $B=7$~T (here at $\sigma_+$), in accordance with the behavior of an exciton that becomes optimal. Second, the trion resonance (here at $\sigma_-$) redshifts in energy without broadening or decay  when the charge density increases, in accordance with the behavior of a distinct resonance of an optimal trion (i.e., when its photoexcited e-h pair becomes distinguishable). These trends are similar to the ones seen in TMD monolayers. 

Finally, Fig.~\ref{fig:QWs}(c) shows the electro-reflectance spectra at progressively increased electron densities in GaAs QWs (bottom to top), taken from Ref.~\cite{Shields_PRB95}. The temperature is  2~K and the magnetic field is $8$~T. The singlet and triplet resonances, marked by $X^-_s$ and $X^-_t$, redshift in energy without broadening when the charge density increases, until they eventually merge at large electron density. Appendix~\ref{sec:caveats} discusses the emergence of a triplet trion in this QW system. According to our proposed framework, an energy redshift without broadening is to be expected if these trions are optimal and their resonances distinct. This is certainly true for the singlet trion, whose electrons are distinguishable (opposite spins in this case). Identifying the triplet resonance in this case as distinct is more subtle because its electrons have the same spin and both reside in the $\Gamma$-valley of the conduction band in GaAs.  Yet, the LL behaves as additional distinguishability knob when the energy difference between LLs exceeds the binding energy of the triplet trion  (1~meV in GaAs QWs). The resonance of the triplet trion under a strong magnetic field is distinct because the photoexcited e-h pair weakly binds to an electron in the lowest LL \cite{Finkelstein_PRB96}, where its large LL degeneracy means that the Pauli principle no longer provides such a strong constraint as at zero field  \cite{Whittaker_PRB97}. In agreement with this understanding, the triplet trion is not observed in these QWs at small magnetic fields \cite{Finkelstein_PRB96,Whittaker_PRB97,Andronikov_PRB05,Homburg_JCG00}.}

\begin{figure*} 
\centering
\includegraphics[width=16cm]{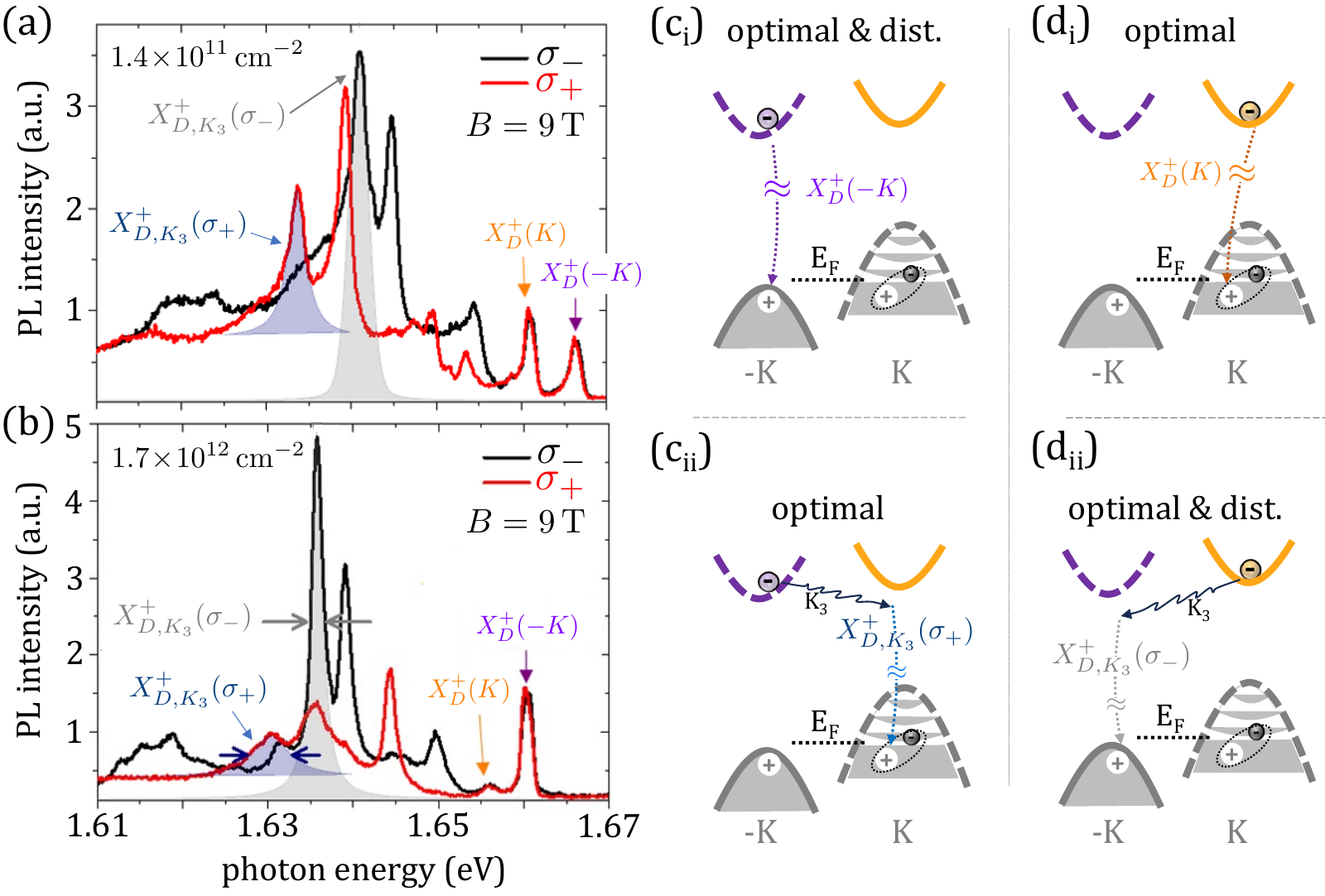}
\caption{Photoluminescence (PL) measured from the spectral window of the dark positive trion in a WSe$_2$ monolayer at $B=9$~T  when the hole density is (a) 1.4$\times$10$^{11}$~cm$^{-2}$ and (b) 1.7$\times$10$^{12}$~cm$^{-2}$. The red (black) line corresponds to detection of circularly polarized emission with $\sigma_+$ ($\sigma_-$) helicity. Taken from Ref.~\cite{Robert_PRL21}. The highlighted resonances are of dark trions, $X^+_D(\pm K)$,  and their $K_3$-phonon replicas, $X^+_{D,K_3}(\sigma_{\pm})$. (c) and (d) Diagrams of the corresponding emission processes (see text). We have omitted the optically-active upper spin-split CB valleys (see Fig.~\ref{fig:WSe2}) since optical transitions of dark trions involve the lower CB valleys.}\label{fig:k3}   
\end{figure*}

\subsection{Ruling out screening by resident carriers as a source of decay of excitonic resonances} \label{sec:screen}

We have suggested that screening can explain the steady energy redshift of distinct resonances of optimal excitonic complexes.  We wish to examine if this screening plays a significant role in the decay and broadening of other optical resonances. For example, the resonances of trions in MoSe$_2$ at $B$$\,$$=\,$$0$, marked by $X^{\pm}$ in Fig.~\ref{fig:dr}(b), represent indistinct resonances of optimal complexes. The same is applicable for the positive trion in WSe$_2$, marked by $X^{+}$ in Fig.~\ref{fig:dr}(a). These resonances behave qualitatively differently than distinct resonances of optimal complexes. When the charge density increases, indistinct resonances of optimal complexes decay and broaden, while showing energy blueshift in absorption at elevated charge densities. We will first address their decay and broadening and in Section~\ref{sec:shakeup} we will focus on their energy shift, which goes beyond the considerations of band filling, BGR, and binding energy. 

To better understand the origin for decay and broadening of indistinct resonances, we compare the behavior of $X^+$ to that of $H_{\mathcal{B}}$ either in Figs.~\ref{fig:dr}(a) or (b). While type-A positive trions are quenched when the gate voltage is approaching ~$-10\,$V, the amplitudes and linewidths of the type-B hexcitons are not affected by increasing the density of holes. Since charge particles screen charge particles, the screening effect cannot be selective in weakening the Coulomb attraction between the particles of one complex species but not of another. Thus, the different behavior of $X^+$ and $H_{\mathcal{B}}$ in either Figs.~\ref{fig:dr}(a) or (b) suggests that screening does not play a significant role in the decay and broadening of indistinct resonances of optimal excitonic complexes (at least in the studied range of charge densities of these experiments). That distinct resonances of optimal excitonic complexes neither decay nor broaden at the \textit{same} charge density that other resonances do broaden, means that the distinguishability of the photoexcited e-h pair and  optimality of the excitonic complex play the important roles in setting their decay and broadening. 

Clear-cut evidence against screening as a source of decay and broadening can be inferred by studying the recombination of dark trions in tungsten-based monolayers. Figure~\ref{fig:k3} shows the photoluminescence (PL) of dark positive trions in a WSe$_2$ monolayer at $B=9$~T when the hole density is (a) 1.4$\times$10$^{11}$~cm$^{-2}$ and (b) 1.7$\times$10$^{12}$~cm$^{-2}$ \cite{Robert_PRL21}. We first explain what is being measured and then we relate the findings to the role of screening. The red (black) curve corresponds to detection of circularly polarized emission with $\sigma_+$ ($\sigma_-$) helicity. Emission due to recombination of dark trions without phonon assistance is denoted by $X^+_D(-K)$ and $X^+_D(K)$ in Figs.~\ref{fig:k3}(a) and (b), and the corresponding  recombination diagrams are shown in Figs.~\ref{fig:k3}(c$_i$) and (d$_i$). The unpolarized light
emission is a result of the out-of-plane dipole orientation \cite{Slobodeniuk_2DMater16}, yielding similar amplitudes for emission with $\sigma_+$ and $\sigma_-$ helicity. Polarized emission is restored when the recombination is assisted by zone-edge phonons, such as the $K_3$ mode, through which the electron experiences an intervalley transition and then recombines with a hole of matching spin in the opposite valley \cite{Dery_PRB15,Yang_PRB22}. This emission is denoted by $X^+_{D,K_3}(\sigma_{\pm})$ in Figs.~\ref{fig:k3}(a) and (b), and the corresponding recombination diagrams are shown in Figs.~\ref{fig:k3}(c$_{ii}$) and (d$_{ii}$).

Coming back to the screening question, we first notice that we are dealing with different recombination processes of the \textit{same} optimal trion. For example, we say that the optical transition in Fig.~\ref{fig:k3}(c$_{i}$) is distinct because the recombining hole belongs to an empty valley (VB valley at $-K$). On the other hand, we can also say that the phonon-assisted optical transition in Fig.~\ref{fig:k3}(c$_{ii}$) of the very same optimal trion is indistinct because now the recombining hole belongs to a populated valley (VB valley at $K$). One cannot argue that  broadening and decay of $X^+_{D,K_3}(\sigma_{+})$ in Fig.~\ref{fig:k3}(b) compared with Fig.~\ref{fig:k3}(a) is caused by screening because the very same trion shows no signs of decay and broadening when measured through its emission from $X^+_D(-K)$. In fact, the latter resonance is stronger and slightly narrower at larger charge density.  

Similarly, we can look at dark positive trions whose electron resides in the opposite valley, as shown in Figs.~\ref{fig:k3}(d$_{i}$) and ~\ref{fig:k3}(d$_{ii}$).  As before, resonances of this optimal trion are said to be distinct or indistinct based on the identity of the recombining hole. Inspecting the resonance $X^+_{D,K_3}(\sigma_{-})$ in Figs.~\ref{fig:k3}(a) and (b), we see that increasing the hole density renders the phonon-assisted emission narrower and stronger, corroborated by the fact that this is a distinct resonance. On the other hand, the indistinct resonance $X^+_{D}(K)$ in Figs.~\ref{fig:k3}(a) and (b)  shows significant decay and broadening when the charge density increases. We therefore cannot conclude that screening is causing this decay, because emission from the very same trion is narrower and stronger if it involves recombination of the distinguishable hole.

 \section{Shakeup Processes} \label{sec:shakeup}

{\color{black} We have seen above that screening is not the source of broadening and energy blueshift of excitonic resonances (moreover, in Section V we will also show that band filling is not a primary source of the energy blueshifts observed in TMD monolayers).  Here, in this section, we elaborate on the shakeup processes that do cause energy shifts and broadening when the e-h pair is indistinguishable and the excitonic complex is optimal.  

The concept of shakeup processes during photoexcitation was introduced originally to understand the Fermi edge problem in metals \cite{Mahan_PR67a, Nozieres_PR69,Schotte_PR69,JP71,Swarts_PRL79}.  In this case, x-ray excitation promotes a deep core electron to the conduction band, leaving behind a hole that remains bound to the atom. The shakeup process is the response of the conduction electrons to the sudden appearance of the localized hole in their vicinity. Around the same time, Mahan predicted a Fermi-edge singularity (FES) in the optical spectra of degenerate semiconductors \cite{Mahan_PR67b}, whose hallmark in the spectrum is an asymmetric resonance that shows energy blueshift and broadening when the charge density increases until the resonance is eventually muted  \cite{Mahan_PR67b,SchmittRink_PRB86,Hawrylak_PRB91}. This prediction was confirmed in semiconductor quantum wells \cite{Skolnick_PRL87}, wherein the photoexcited hole was localized by defects or interface fluctuations, rendering its mass ‘infinite’. 


To date, investigations that studied the relation between FESs and shakeup processes in degenerate semiconductors \cite{Chang_PRB85,Sooryakumar_PRL87,Kane_PRB94,Finkelstein_PRB97,Mkhitaryan_PRL11}, have overlooked the key role played by optimality and distinguishability. We have seen that while certain absorption resonances in TMD monolayers show FES-like behavior (blueshift and broadening), the behavior of other resonances in the same monolayer and in the presence of the same Fermi sea is qualitatively different (no blueshift and no broadening). This inconsistency indicates that the understanding of shakeup processes and the conditions at which they emerge have to be refined.

To help us introduce shakeup processes that take into account the optimality and distinguishability, we first explain the relation of these processes to scattering through the lens of Anderson's infrared catastrophe \cite{Anderson_PRL67}. Adjusting his model to our problem, we consider $N$ resident electrons in a monolayer and assume that their density is large enough, such that the system is effectively a noninteracting Fermi gas. Upon sudden appearance of a short-range potential  (e.g., a photoexcited e-h pair), the overlap between Slater determinants of the $N$-electron systems before and after the perturbation becomes \cite{Anderson_PRL67}
\begin{equation}
S \sim N^{\displaystyle{- \alpha \sin^2 \delta}},
\end{equation}
where $\alpha$ is a constant of the order of 1, and $\delta$ is the effective phase shift of Fermi surface electrons because of their scattering with the short-range potential. The infrared catastrophe alluded by Anderson refers to cases where the Slater determinants are orthogonal ($S \rightarrow 0$ when $\delta \neq 0$), meaning that the sudden introduction of the short range potential cannot exist without a shakeup of the Fermi surface electrons.  This process involves low-energy excitations of resident electrons to states above the Fermi surface, which can manifest by emission of collective excitations such as plasmons or magnons. On the other hand, a sudden perturbation without this shakeup can be realized if $S \rightarrow 1$ (i.e., $\delta \rightarrow 0$). 

\begin{figure*}[t!]
\centering
\includegraphics[width=16cm]{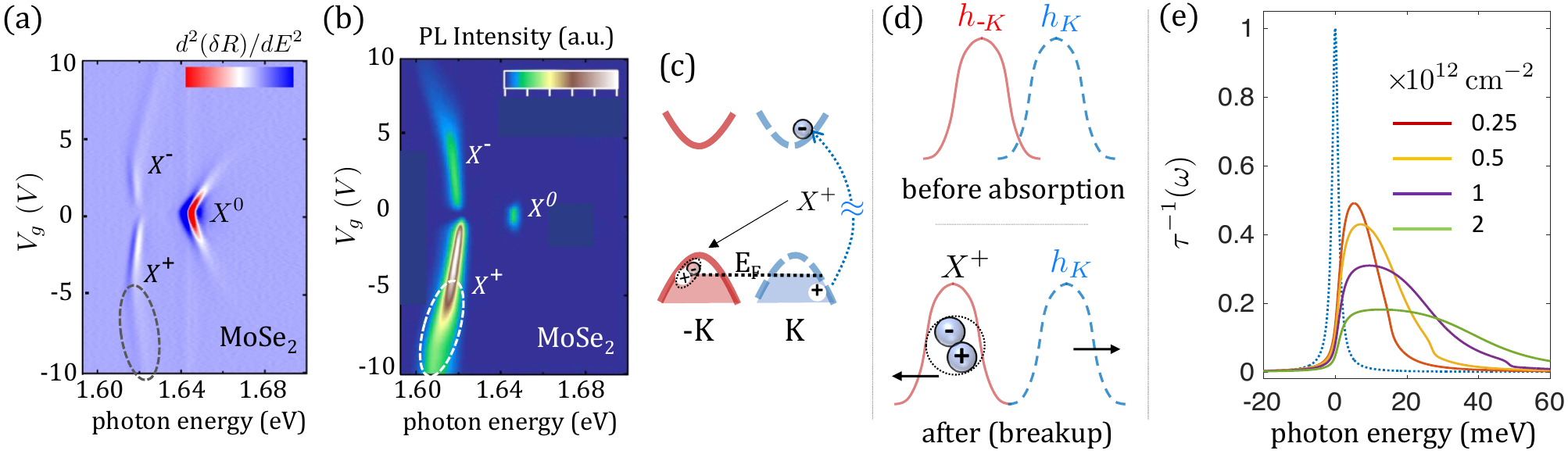}
\caption{Colormaps of (a) reflectance and (b) photoluminescence spectra in MoSe$_2$ at $B=0$. Taken from Ref.~\cite{Liu_NatComm21}. The indistinct resonances of the optimal trion complexes, labeled by $X^{\pm}$, broaden when resident carriers are added to the monolayer (readily seen when $|V_g| \gtrsim4$~V). (c) Diagram of the positive trion following photoexcitation of the $K$ valley. (d) Breakup process following the creation of $X^+$. {\color{black}(e) Calculated rates of trion absorption due to a shakeup by plasmons in hole-doped monolayer MoSe$_2$ (solid lines). The calculated rates are normalized with respect to the rate of the direct trion absorption  in the low density limit (dotted line). The latter corresponds to absorption without plasmon assistance, where the photon energies are measured with respect to its resonance.}
}\label{fig:breakup} 
\end{figure*}

In electrostatically-doped semiconductors, $S \rightarrow 1$ amounts to zero-line optical transitions (i.e., without shakeup of the Fermi sea), described by distinct resonances of optimal excitonic complexes. We explain this scenario through photoexcitation of a negative trion in MoSe$_2$ monolayer whose resident electrons are fully valley polarized, as shown in Fig.~\ref{fig:MoSe2}(f). When the distinguishable e-h pair binds to a resident electron at position $\mathbf{r}$, the Pauli exclusion principle means that no other resident electrons were present at $\mathbf{r}$ and that no other electrons can approach $\mathbf{r}$. Namely, the creation of this trion does not require scattering of the other resident electrons (zero phase shift, $\delta=0$). From a wave-like perspective, the trion inherits the wavevector of the electron and keeps the same crystal momentum after photoexcitation. In other words, the photoexcited e-h pair `catches a ride' with the resident electron to which it binds, so that the Pauli exclusion principle with other resident electrons is not violated. This process describes a different physics than the one provided by the Fermi polaron theory, wherein an attractive polaron is formed when Fermi sea particles approach a photoexcited e-h pair  \cite{Sidler_NatPhys17,Efimkin_PRB17}.  In the Fermi polaron picture, the resident carriers respond by moving towards the e-h pair, and this motion mandates low-energy excitations across the Fermi surface (i.e., $\delta \neq 0$), which in turn leads to scattering and  broadening. However, we have seen that regardless of the semiconductor type, a distinct resonance of an optimal excitonic complex does not broaden when the charge density increases.

Below, we will focus on shakeup processes that accompany the optical transition in the opposite extreme, $S \rightarrow 0$ ($\delta \neq 0$). Applicable to indistinct resonances of optimal complexes, the shakeup  involves a breakup (i.e., physical separation) of distinguishable and indistinguishable resident carriers upon photoexcitation, or their reunion after recombination. These processes will help us to self-consistently explain the experimental results.}  

\subsection*{Energy shifts and broadening of indistinct resonances of optimal complexes through breakup and reunion processes} \label{sec:breakup}

When distinguishable and indistinguishable resident carriers are present, breakup or reunion processes govern the behavior of indistinct resonances of optimal complexes. We first explain these processes in charge tunable MoSe$_2$ monolayers at $B$=0, due to its relatively simple absorption and emission spectra, shown in Figs.~\ref{fig:breakup}(a) and (b) \cite{Liu_NatComm21}.  In absorption (emission), the indistinct resonances of optimal trions blueshift (redshift) and broaden in energy. Without loss of generality, we analyze these phenomena by focusing on the indistinct resonance $X^+$ whose absorption process in the $K$ valley is shown in Fig.~\ref{fig:breakup}(c). 

Prior to the absorption of $X^+$, resident holes from the VB valleys at $-K$ and $K$ have some spatial overlap. This partial overlap is illustrated through the wave-packet envelopes of the holes in the top diagram of Fig.~\ref{fig:breakup}(d). Mandated by the Pauli exclusion principle, this overlap breaks when the resident hole from $-K$ binds to the photoexcited e-h pair from $K$ to form $X^+$, illustrated by the bottom diagram of Fig.~\ref{fig:breakup}(d), where the arrow signifies that resident holes from $K$ are forced out of the trion's region. The energy of the absorbed photon is converted to $ \hbar\omega = \varepsilon_T + \delta_\varepsilon$, where $\varepsilon_T$ is the trion energy and $\delta_\varepsilon$ is the energy required to break the overlap, through which we can understand the energy blueshift.  Since the overlap is larger with increasing charge density and since it costs more energy to break a stronger overlap, the energy blueshift  scales with charge density. Furthermore, the energy blueshift of $X^+$ in Fig.~\ref{fig:breakup}(a) indicates that $\delta_\varepsilon$ is stronger than the underlying redshift effect from screening at elevated densities (Sec.~\ref{sec:bgr}). {\color{black} To quantify the spectral broadening due to this shakeup process, we consider the plasmon-assisted absorption of a trion, where plasmon emission creates a charge wave that facilitates the ousting of indistinguishable holes from the trion creation region. Figure~\ref{fig:breakup}(e) shows the resulting absorption spectrum, where technical details of this trion-plasmon process are provided in Appendix~\ref{sec:trion_plasmon}.  The broadening of $X^+$ reflects more than one way to shakeup the system, where photon absorption at a larger energy involves emission of a plasmon with shorter wavelength.}

\begin{figure}
\centering
\includegraphics[width=8.5cm]{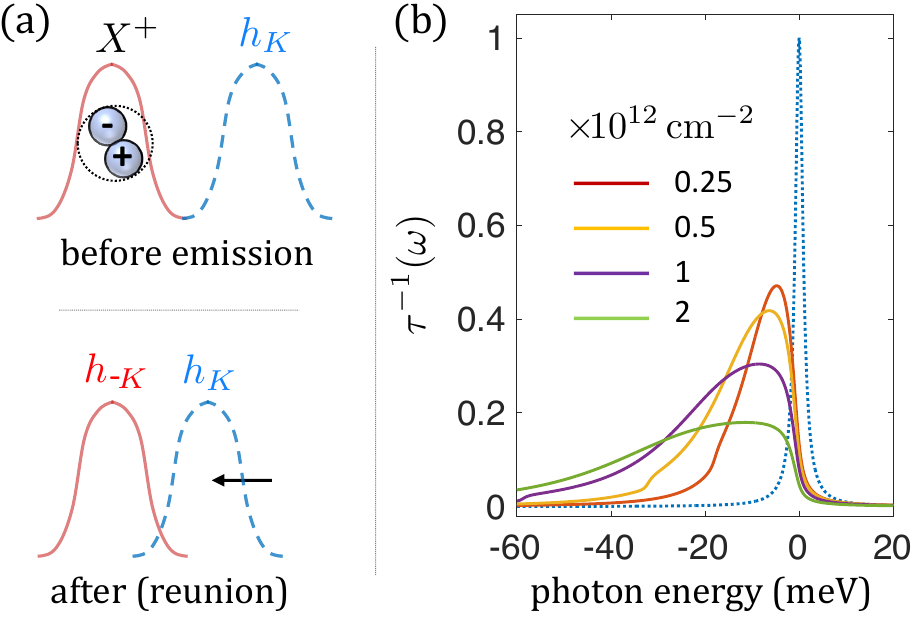}
\caption{ {\color{black} (a) Shakeup (reunion) process following the recombination of $X^+$. (b) Calculated rates of trion emission due to a shakeup by plasmons in hole-doped monolayer MoSe$_2$ (solid lines). The calculated rates are normalized with respect to the rate of the direct trion emission in the low density limit (dotted line). The latter corresponds to emission without plasmon assistance, where the photon energies are measured with respect to its resonance.}
}\label{fig:reunion} 
\end{figure}

Figure~\ref{fig:reunion} shows the respective process during recombination. Mandated by the Pauli exclusion principle, the top diagram in Figure~\ref{fig:reunion}(a) shows the lack of overlap between the wave-packet envelopes of a hole from $K$ and a trion with indistinguishable e-h pair from $K$. After emission, the overlap between the left-behind hole from $-K$ and a resident hole from $K$ is restored, as shown by their reunion in the bottom diagram of Fig.~\ref{fig:reunion}(a). The recombination process leaves behind an inhomogeneous Fermi sea because the region that was occupied by the trion is left void of indistinguishable resident carriers, which are then pushed back to replenish this region as indicated by the arrow in Fig.~\ref{fig:reunion}(a). The energy of the trion before recombination is converted to $\varepsilon_T = \hbar\omega + \delta_\varepsilon$, where $\delta_\varepsilon$ in this case is the energy difference between the inhomogeneous Fermi sea immediately after recombination and the homogeneous one in equilibrium. Since $\delta_\varepsilon$ is commensurate with the hole density, the conversion leaves the emitted photon with less energy when the charge density increases, in line with the measured energy redshift inside the highlighted dashed white oval in Fig.~\ref{fig:breakup}(b). The redshift is further enhanced by the interplay between BGR and reduced binding energy due to screening (Sec.~\ref{sec:bgr}).  {\color{black}Figure~\ref{fig:reunion}(b) shows the calculated recombination rate when considering trion emission that is accompanied by plasmon emission (Appendix~\ref{sec:trion_plasmon}). The plasmon in this case pushes indistinguishable resident carriers into the region that was left void after recombination, thereby restoring the homogeneity of the Fermi sea}.

\begin{figure}
\centering
\includegraphics[width=8.5cm]{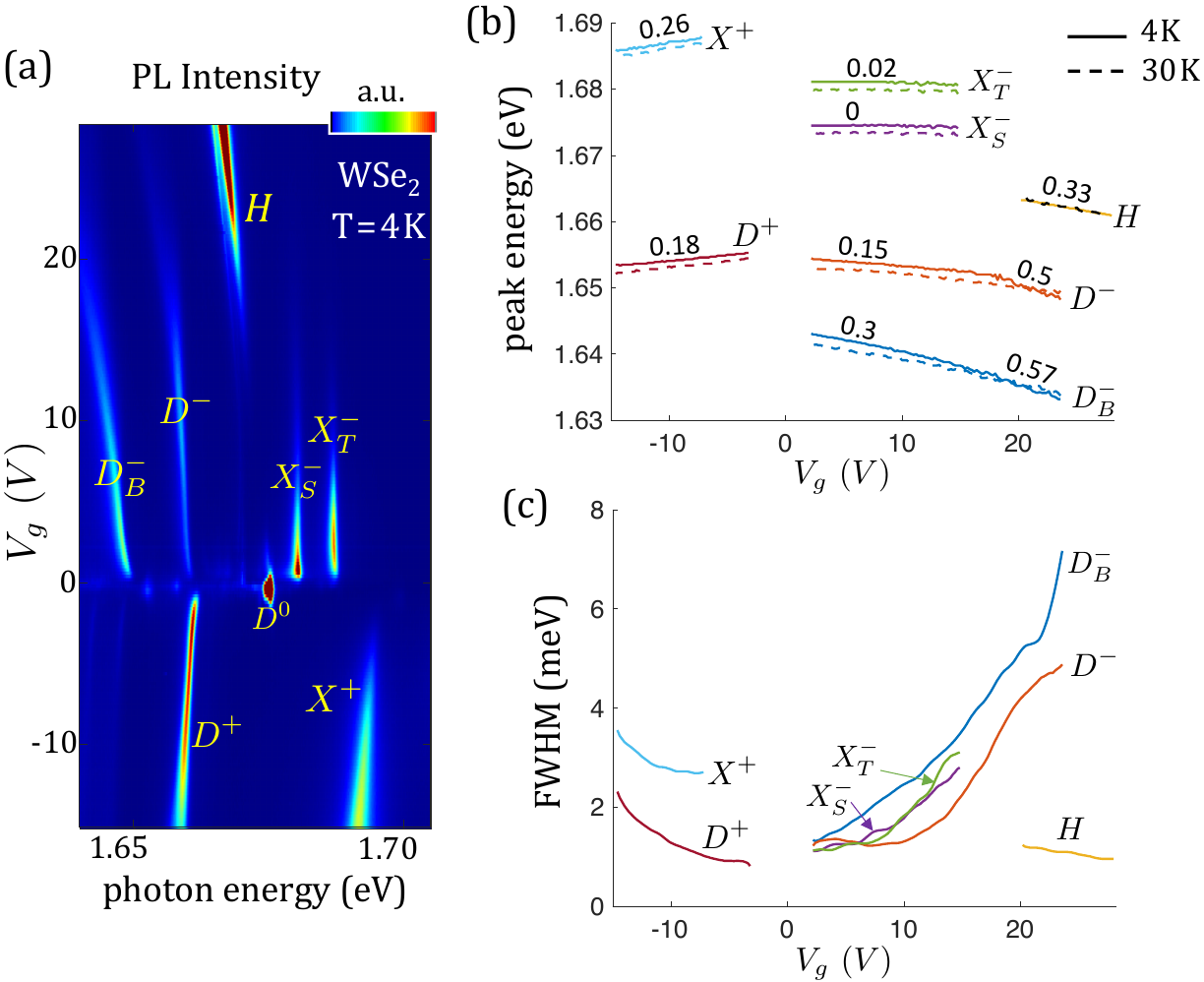}
\caption{Photoluminescence intensity colormap of WSe$_2$ at $B\,$$=\,$0, shown in the spectral window of trions. The laser energy is 1.96~eV and its power is 1~$\mu$W. (b) and (c) Extracted peak energies and full-width at half maximum of the resonances in (a). The numbers in (b) denote the energy redshift in meV per 10$^{11}$$\,$cm$^{-2}$, where a change of 1$\,$V in this device corresponds to a change of 10$^{11}$$\,$cm$^{-2}$ in charge density. }\label{fig:wse2case} 
\end{figure}

The self-consistency of the breakup and reunion shakeups is examined through the unique case of WSe$_2$ monolayers. Figure~\ref{fig:wse2case}(a) shows the spectral window of trion emission in a charge tunable WSe$_2$ monolayer at T$\,$$=$$\,$4$\,$K. Details of the sample fabrication and experimental setup can be found in the supplementary information of Ref.~\cite{Robert_NatComm21}, and analysis of the dark exciton and trion resonances, $D^0,\,D^+,\,D^-,\,D^-_B$ can be found in Ref.~\cite{Yang_PRB22}.  Figures~\ref{fig:wse2case}(b) and (c) show the extracted resonance energies and their full width at half maximum (FWHM), respectively, where the dashed lines show the resonances energies at T$\,$$=$$\,$30$\,$K as well. The numbers in Fig.~\ref{fig:wse2case}(b) denote the energy redshift of the corresponding resonance in meV per 10$^{11}$$\,$cm$^{-2}$. The energy redshift of $H$ and its lack of broadening, as summarized in Figs.~\ref{fig:wse2case}(b) and (c), are hallmarks of the distinct resonance of the optimal hexciton. The resonance does not broaden since the Fermi sea is not perturbed in this case, and as discussed in Sec.~\ref{sec:bgr}, its energy redshift stems from the interplay between BGR and reduced binding energy due to screening. 

The energy redshift of $X^+$ in both MoSe$_2$ and WSe$_2$, evidenced from the emission spectra in Figs.~\ref{fig:breakup}(b) and ~\ref{fig:wse2case}(a), comes from the fact that their positive trions are optimal complexes with indistinct resonances. Namely, the emission of $X^{+}$ is accompanied by a reunion process. The enhanced broadening of $X^{\pm}$ in Fig.~\ref{fig:breakup}(b) compared with that of $X^+$ in Fig.~\ref{fig:wse2case}(a) is attributed to the larger densities used in the MoSe$_2$ device, where $-$10~V in Fig.~\ref{fig:wse2case}(a) is the equivalent of $-$1.5~V in Fig.~\ref{fig:breakup}(b). Because recombination of an optimal dark trion at $B=0$ is governed by optical transition of an indistinguishable e-h pair, reunion processes are also the reason that dark trion resonances broaden and redshift in energy, as shown in Fig.~\ref{fig:wse2case}. Finally, of all resonances in the emission spectra of Figs.~\ref{fig:breakup}(b) and ~\ref{fig:wse2case}(a), the only ones that do not shift in energy are the triplet and singlet negative trions ($X^-_{S,T}$). As we show next, this behavior is consistent with the fact that $X^-_{S,T}$ are distinct rather than indistinct resonances. 

\begin{figure}[t]
\centering
\includegraphics[width=8.5cm]{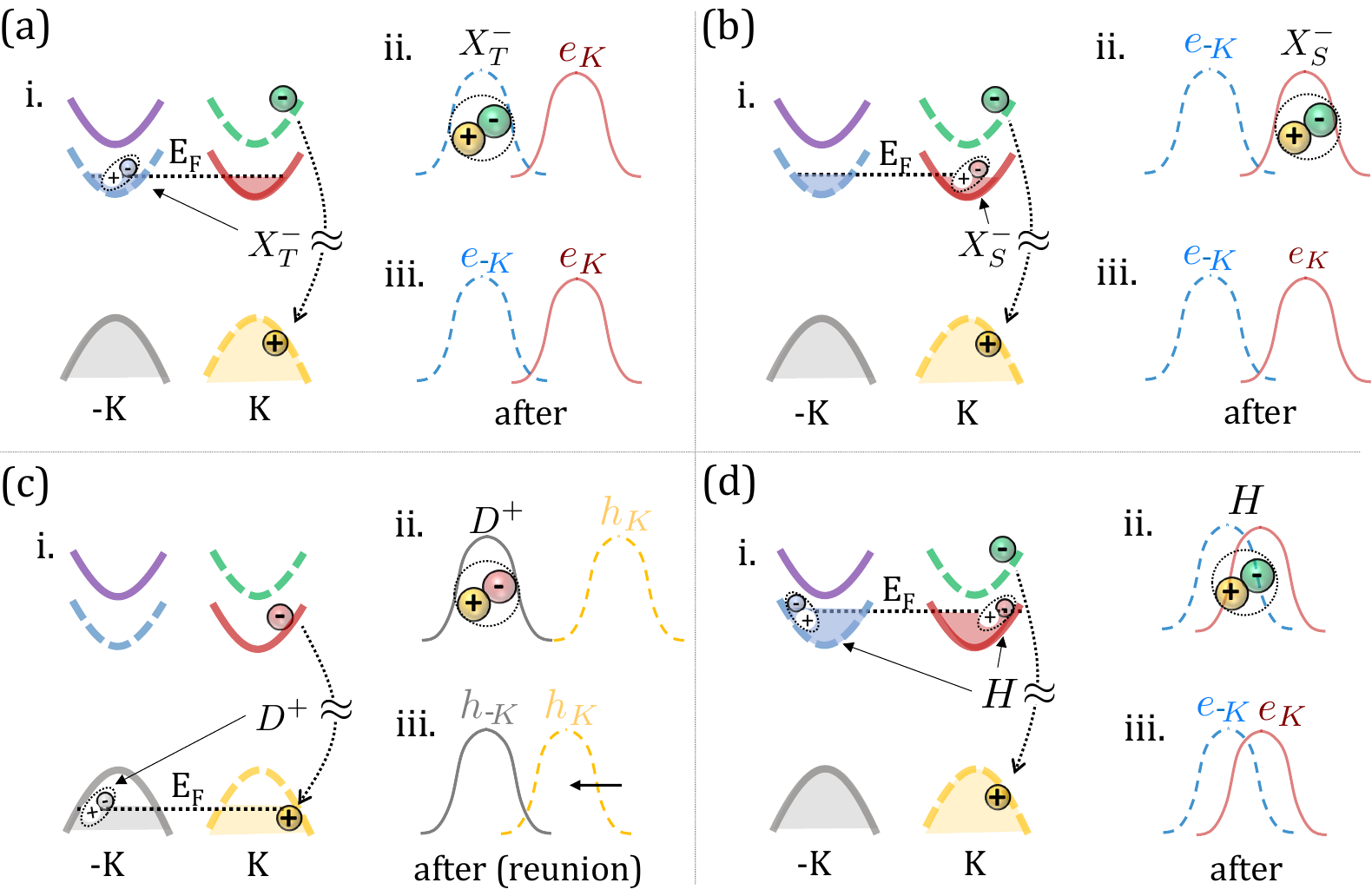}
\caption{ Diagrams of the triplet and singlet negative trions, positive dark trion and hexciton in WSe$_2$ monolayer. For each complex, (i) corresponds to its energy diagram, and (ii)/(iii) is the corresponding real-space configuration before/after recombination. Of the 4 shown complexes, only $D^+$ involves recombination of an indistinguishable e-h pair, and therefore its emission is followed by a reunion process.}\label{fig:wse2schemes} 
\end{figure}

The energy diagrams of the triplet, singlet, dark positive trion, and hexciton in WSe$_2$ monolayer are illustrated in parts (i) of Figs.~\ref{fig:wse2schemes}(a)-(d). The use of different colors is meant to distinguish between particles with  different spin and valley quantum numbers. As shown in Fig.~\ref{fig:wse2schemes}(a), there is no restriction for an electron from the bottom CB valley at $K$ to be completely excluded from the immediate vicinity of the triplet trion, illustrated by the slight overlap of $X^-_{T}$ and $e_K$ in diagram (ii). The same is applicable for an electron from the bottom CB valley at $-K$ and the singlet trion as shown by the corresponding diagram in Fig.~\ref{fig:wse2schemes}(b). After recombination of the trion, the electron that is left behind  is already in proximity with an electron from the time-reversed valley, as shown by diagrams (iii) of Figs.~\ref{fig:wse2schemes}(a) and (b). That is, the distribution of resident carriers remains homogeneous in the presence of $X^-_{S,T}$. The result is that the distinct resonances of $X^-_{S,T}$ do not redshift in energy because there is no need for reunion processes  to follow the recombination. Contrary to these cases, other trion species require reunion processes because the Pauli exclusion principle is at play, as illustrated in Fig.~\ref{fig:wse2schemes}(c) for the positive dark trion. Finally, Fig.~\ref{fig:wse2case}(a) shows that when the electron density is larger than 10$^{12}$~cm$^{-2}$ ($V_g > 10\,$V), $X^-_{S,T}$ decay and $H$ emerges. The corresponding hexciton diagrams are shown in Fig.~\ref{fig:wse2schemes}(d), where the e-h pair simultaneously  binds to two nearby electrons from the bottom CB valleys at $K$ and $-K$. 

Comparing the behavior of $X^-_{S,T}$ to that of $H$, we should explain why $X^-_{S,T}$ are not subjected to the same screening-induced energy redshift like $H$, and what causes the broadening of $X^-_{S,T}$. The broadening indicates a shorter lifetime when the electron density increases, caused by the trion-electron interaction which renders them dark through exchange scattering (i.e., $X^-_{S,T} \rightarrow D^-$) \cite{Yang_PRB22}. The lack of screening-induced energy redshift, on the other hand, is less clear. We can only conjecture that if the resident electrons are quasi-localized when the electron density $n_e$ is a few times $10^{11}$$\,$cm$^{-2}$ or less, then $X^-_{S,T}$ in Fig.~\ref{fig:wse2schemes} are not subjected to the screening-induced energy redshift that $H$ is subjected to when the resident electrons become itinerant at $n_e > 10^{12}$$\,$cm$^{-2}$ (Appendix \ref{sec:localization}). 

\begin{figure*}[t!]
\centering
\includegraphics[width=16cm]{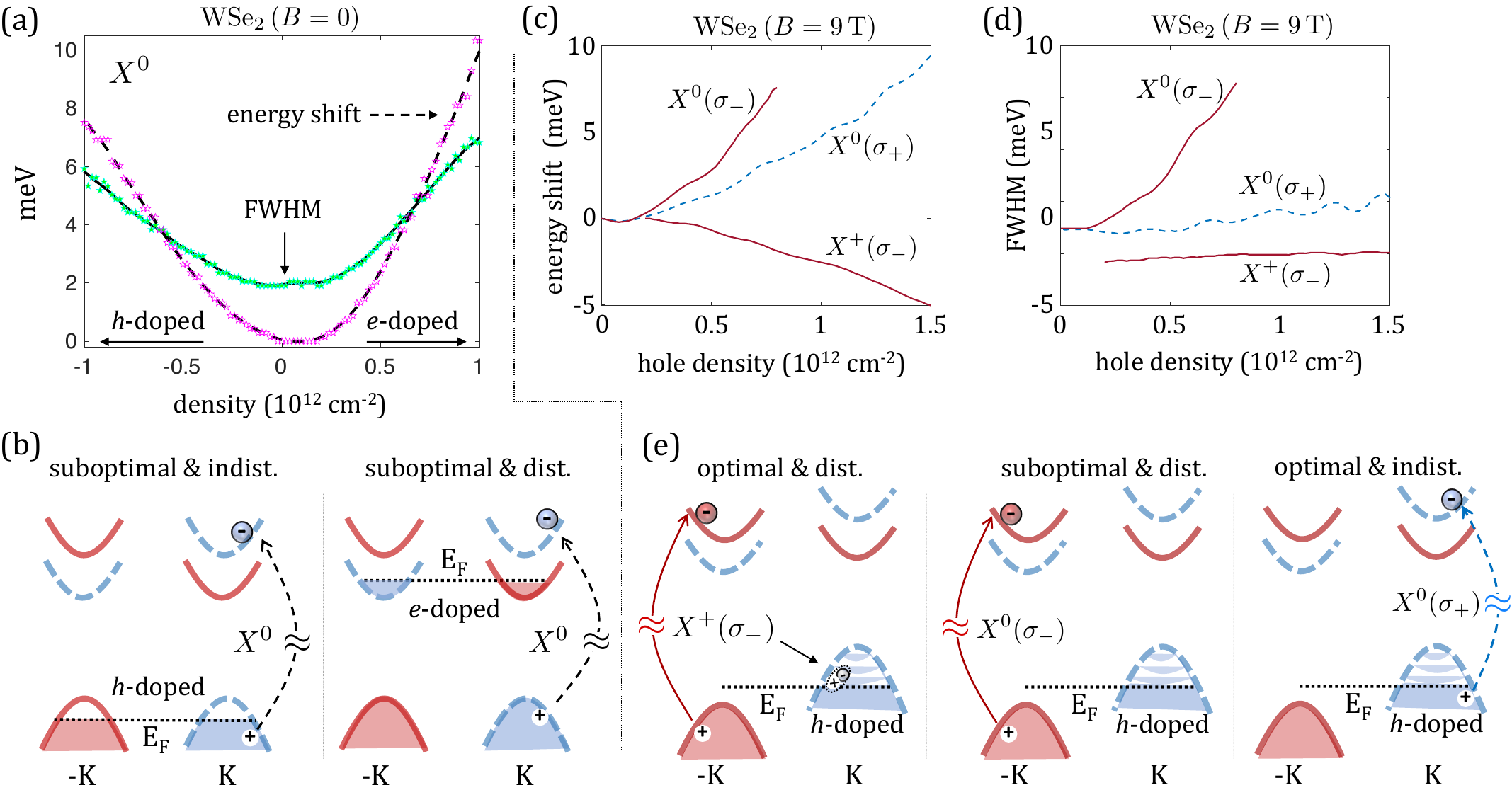}
\caption{(a) Energy shift and FWHM of the neutral exciton resonance as a function of charge density, extracted from low-temperature reflectance spectra of hBN-encapsulated WSe$_2$ monolayer. (b) The absorption processes in the hole- and electron-doped regimes are shown in the left and right diagrams, respectively. (c)-(e) Corresponding helicity-resolved results in hole-doped monolayer at $B=9\,$T.  Also shown are results of the distinct resonance of the optimal positive trion, labeled by $X^+(\sigma_-)$. The energy blueshift and FWHM of $X^0(\sigma_-)$ are shown in a range of hole densities at which this resonance is still identifiable in the reflectance spectra (its oscillator strength is rapidly depleted in favor of $X^+(\sigma_-)$).}\label{fig:1s} 
\end{figure*}
{\color{black}

\section{Energy shift of neutral excitons} \label{sec:blueX0}

We have identified in Sec.~\ref{sec:obs} that distinct resonances of optimal excitonic complexes redshift in energy when distinguishable resident carriers are added to the system. In addition, we have identified in Sec.~\ref{sec:shakeup} that indistinct resonances of optimal excitonic complexes tend to shift and broaden in energy when indistinguishable resident carriers are added to the system. Another interesting phenomenon that we will discuss in this section is that the energy blueshift of the neutral-exciton absorption resonance is relatively weak when the exciton is optimal (i.e., in the absence of distinguishable resident carriers). This result is counterintuitive because adding distinguishable resident carriers does not Pauli block the optical transition, whereas adding indistinguishable resident carriers leads to energy blueshift from the Moss–Burstein effect  -- namely, the optical energy gap increases when indistinguishable resident carriers occupy the low energy states in the photoexcited valley up to the Fermi energy. 

We demonstrate and quantify the energy blueshift of the absorption resonances of neutral excitons in WSe$_2$ monolayers.} The choice to focus on absorption resonances in this monolayer is intentional because we can then readily decipher the roles of distinguishability and optimality in three possible configurations of the exciton under various conditions. On the other hand, the recombination and energy relaxation of neutral excitons involve exchange scattering processes that complicate the analysis of emission resonances \cite{Yang_PRB22}. Exchange scattering is  also relevant when dealing with absorption resonances of excited-state neutral excitons, and we leave the analysis of this case to Appendix~\ref{sec:exc_scat}. 

Figure~\ref{fig:1s}(a) shows the energy shift and FWHM of the neutral exciton absorption resonance as a function of charge density in an hBN-encapsulated WSe$_2$ monolayer at zero magnetic field. These results  were extracted from low-temperature reflectance spectra, where details of the sample fabrication and experimental setup can be found in the supplementary information of Ref.~\cite{Robert_NatComm21}. When the hole density increases from zero to 10$^{12}$~cm$^{-2}$, the energy blueshift is 7.5~meV and the FWHM increases from 2 to 6~meV. Half of the resident holes are distinguishable and the other half are indistinguishable, as shown by the left diagram of Fig.~\ref{fig:1s}(b). When the monolayer is electron-doped, the corresponding energy blueshift is 10~meV and the FWHM increases from 2 to 7~meV.  The resident electrons are 100\% distinguishable in this case, as shown by the right diagram of Fig.~\ref{fig:1s}(b). Assuming that distinguishability of the resident carriers is the key point regardless of whether these are electrons or holes (as long as their effective masses are comparable), we can use the results of Fig.~\ref{fig:1s}(a) to infer the following. Increasing the density of distinguishable (indistinguishable) resident carriers by 10$^{11}$~cm$^{-2}$ roughly adds 1~meV (0.5~meV) to the blueshift and 0.5~meV (0.3~meV) to the broadening of the neutral exciton resonance. That is, the effect from distinguishable resident carrier is about as twice as strong.

The stronger effect caused by the presence of distinguishable resident carriers is corroborated by helicity-resolved measurements of the energy shift and FWHM from the same device. Figures~\ref{fig:1s}(c) and (d) show the respective results when $B=9\,$T, where diagrams of the corresponding absorption resonances are shown in Fig.~\ref{fig:1s}(e). We can confirm that the holes are fully valley polarized through the steady energy redshift and constant FWHM of the distinct resonance of the positive trion, labeled by $X^+(\sigma_-)$. Focusing on the neutral exciton diagrams in Fig.~\ref{fig:1s}(e), the valley-polarized resident holes are 100\% distinguishable when dealing with the distinct resonance of the suboptimal exciton, labeled by $X^0(\sigma_-)$, and 100\% indistinguishable when dealing with the indistinct resonance of the optimal exciton, labeled by $X^0(\sigma_+)$. Comparing the energy blueshifts of $X^0(\sigma_\pm)$ in Fig.~\ref{fig:1s}(c), we again notice that distinguishable resident carriers induce as twice as strong energy blueshift. The FWHM results in Fig.~\ref{fig:1s}(d) show that the broadening rate of $X^0(\sigma_+)$ is strongly suppressed compared with that of  $X^0(\sigma_-)$. Given that there are no distinguishable resident carriers with which an optimal exciton can form a trion, the broadening of $X^0(\sigma_+)$ is not limited by the same ultrafast process that converts suboptimal excitons to optimal trions (i.e., $X^0(\sigma_-)\,\rightarrow X^+(\sigma_-)$). These results are consistent with previous findings \cite{VanTuan_PRB19,Liu_PRL20}.

{\color{black}

\subsection{Indistinguishable resident carriers} \label{sec:blueX0indis}

We first explain the energy blueshift due to the presence of indistinguishable resident carriers. While the origin of the blueshift is the energy increase of the optical gap, the microscopic details that facilitate the formation of a \textit{bound} exciton are subtle and merit discussion.  Ignoring the presence of distinguishable resident carriers (to be addressed later), we consider the optical excitation scenario in Fig.~\ref{fig:bluex0}(a). Since the absorbed photon carries negligible momentum, creating a bound exciton in the light cone means that its center-of-mass wavevector overlaps with that of a resident electron from the bottom of the valley ($K_x = k_e \sim 0$). Assuming a point-like exciton (e.g., 1$s$), this means that the Pauli exclusion between the electron of the exciton and the resident electron is violated. One way to alleviate this problem is by accompanying the optical excitation with scattering of a resident electron with the photoexcited pair, as shown in Fig.~\ref{fig:bluex0}(b), such that the exciton ends up with $K_x \geq k_F$. 

The dominant role of the exchange interaction between an exciton and resident carriers in semiconductors can facilitate such optical process, wherein the electron (hole) component of the exciton is exchanged with a resident electron (hole) after their interaction \cite{VanTuan_PRB25,Yang_PRB22}. We elaborate on this mechanism in Appendix~\ref{sec:exc_x0}, where we also show that the exchange scattering of the exciton with resident carriers is far stronger than the exciton-phonon interaction or than the direct interaction, wherein the components of the exciton are kept the same before and after scattering. Figure~\ref{fig:bluex0}(c) shows the calculated absorption rate when the exciton creation is accompanied by exciton-hole exchange interaction (see Appendix~\ref{sec:exc_x0} for details). The energy blueshift and broadening match the trends seen in experiment, wherein absorption at a larger energy involves larger momentum transfer during the exciton-hole exchange interaction.

\begin{figure}[t]
\centering
\includegraphics[width=8.5cm]{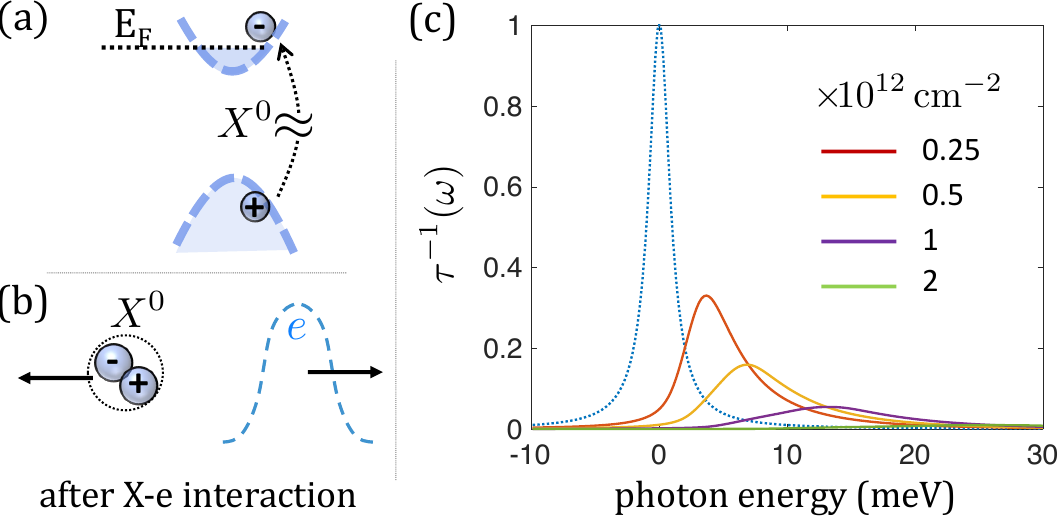}
\caption{{\color{black}(a) Absorption diagram of a suboptimal neutral exciton with indistinct resonance. (b) The optical process is facilitated by the exciton-electron exchange interaction (see text and Appendix ~\ref{sec:exc_x0}). (c)  Rates of exchange-assisted exciton absorption in hBN-encapsulated WSe$_2$ monolayer, calculated in the presence of  indistinguishable resident holes at four densities (solid lines). The calculated rates are normalized with respect to the direct exciton absorption rate at zero charge density (dotted line), whose resonance is the reference energy.}}\label{fig:bluex0} 
\end{figure}

\subsection{Distinguishable resident carriers} \label{sec:blueX0dis}

Band filling by distinguishable resident carriers does not Pauli block the optical transition, and their presence does not Pauli exclude the photoexcited e-h pair. Yet, we have seen that the energy blueshift of suboptimal excitons is strongest when the semiconductor only hosts distinguishable resident carriers (Fig.~\ref{fig:1s}). The presence of these carriers means that there is an optimal complex at a lower energy than the suboptimal exciton. In connection with the scattering phenomenon we have discussed before, the suboptimal exciton refers to a case that the photoexcited e-h pair scatters off a distinguishable resident carrier rather than binds to it.  To study the energy blueshift effect of the suboptimal exciton and the continuous transfer of its oscillator strength to the optimal complex at lower energy, we will render the relation between the scattering phase shift and excitonic density of states (DOS). This relation allows one to understand the relatively strong energy blueshift  of the suboptimal complex by quantifying how the DOS function of the exciton-electron system evolves when the density of distinguishable resident carriers increases.

\begin{figure*}[t!]
\centering
\includegraphics[width=16cm]{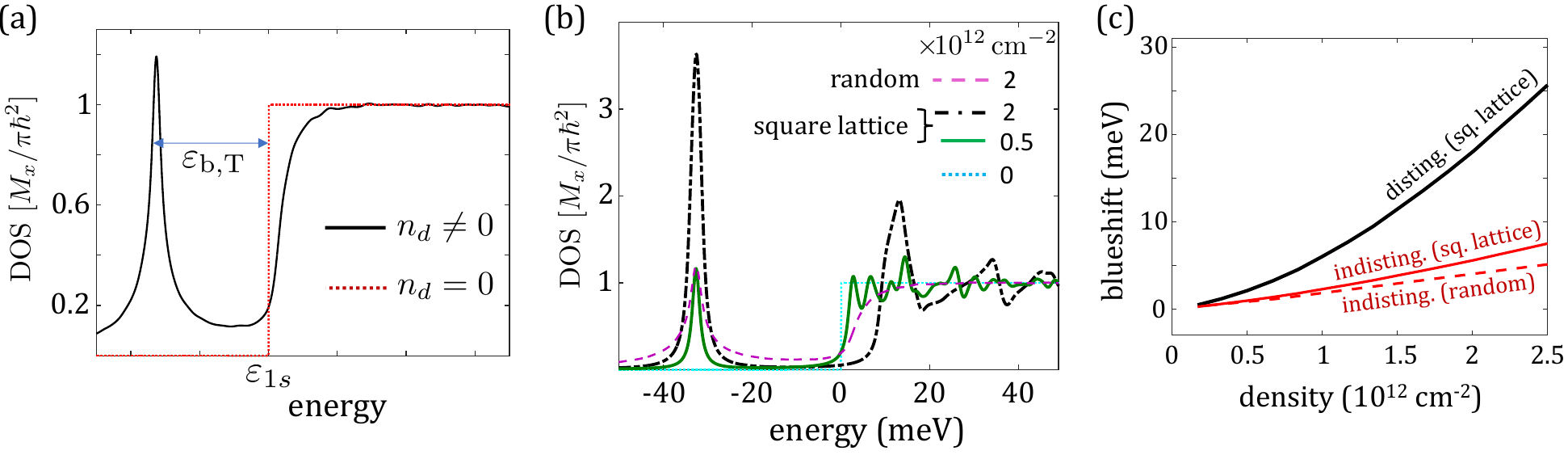}
\caption{{\color{black}(a) Excitonic density of states (DOS) function in a two-dimensional semiconductor with and without distinguishable resident carriers. $\varepsilon_{\text{1s}}$ and $\varepsilon_{\text{b,T}}$ are the resonance energy of the suboptimal exciton and binding energy of the optimal trion, respectively. (b) Calculated excitonic DOS function in electron-doped WSe$_2$ when the electrons form a square lattice and their charge density is $5\times10^{11}$~cm$^{-2}$ (solid green line) and $2\times10^{12}$~cm$^{-2}$ (dashed-dotted black line). The latter is also calculated when the distinguishable electrons are randomly distributed in the sample (dashed magenta line). The dotted cyan line shows the excitonic DOS function without resident electrons, where $\varepsilon_{\text{1s}}$ is the zero reference energy. (c) Calculated density dependence of the energy blueshift of excitons when the distinguishable electrons form a square Wigner lattice (black solid line), indistinguishable and form a square Wigner lattice (red solid line), or indistinguishable and randomly distributed (red dashed line). Taken from Ref.~\cite{VanTuan_PRB23}.}}\label{fig:dos} 
\end{figure*}

Let us consider an effective two-body system comprising a distinguishable resident (itinerant) electron and a point-like exciton. The relative motion between the exciton and electron is governed by a dipole-charge interaction, which if approximated by a central potential, the change to the DOS function for each partial wave can be written as \cite{Gao_PRA17},
\begin{equation}
\Delta D_{l}(\varepsilon)  = \begin{cases}
  \sum_j \delta(\varepsilon-\varepsilon_{j}) \,\,,&  \varepsilon  < 0 \\ \\
\mathlarger{{\frac{1}{\pi} \frac{\delta_l}{d\varepsilon}}} \,\,,& \varepsilon > 0\,.
\end{cases}
\end{equation}
The sum in the first line runs over bound states (i.e., trion states), and the second line corresponds to continuum states (the exciton and resident carrier are not bound to each other). The phase shift $\delta_l$ is introduced by the scattering between the exciton and the electron in this case. Focusing on the optically active $s$ states ($l=0$), the Levinson's theorem implies that \cite{Gao_PRA17,Levinson_MFM49,Newton_book82,Ma_JPMG06}  
\begin{equation}
\int_{-\infty}^{+\infty} \Delta D_{s}(\varepsilon) d\varepsilon = 0\,.
\end{equation}
This is essentially a zero sum rule which states that adding to the DOS in one spectral region mandates taking away from the DOS in a different spectral region. Relevant to the energy blueshift of the suboptimal exciton resonance, forming a bound trion state at $\varepsilon <0$ implies depleting the low-energy DOS region of the exciton at $\varepsilon > 0$.

Figure~\ref{fig:dos}(a) shows characteristic schemes of the DOS function with (solid black line) and without (dotted red line) distinguishable resident carriers. The DOS function units are $M_x/\pi \hbar^2$, where $M_x$ is the translational mass of the exciton in a two-dimensional semiconductor. The exciton resonance energy is $\varepsilon_{\text{1s}}$ and the binding energy of the trion is $\varepsilon_{\text{b,T}}$. The transfer of spectral weight from the low energy part of the step function (slightly above $\varepsilon_{\text{1s}}$) to the trion resonance depends on the details of the scattering phase shift. This transfer gives rise to the energy blueshift of the exciton and to the transfer of oscillator strength from the exciton to the trion \cite{VanTuan_PRB23}. Such transfer does not arise when the semiconductor is only populated by indistinguishable resident carriers, in which case the exciton is optimal and a trion state cannot be created.

The energy blueshift effect due to the transfer of DOS weight from the exciton to the trion spectral regions is a general effect that does not depend on the itinerancy of resident carriers. In case of localized and distinguishable resident carriers, we have used the recursion method to calculate the excitonic DOS function at various charge densities and order configurations. Technical details of these calculations can be found in Ref.~\cite{VanTuan_PRB23}, and here we present the final results. Figure~\ref{fig:dos}(b) shows the excitonic DOS function of electron-doped WSe$_2$ when the electrons are distinguishable and they form a square lattice. Their charge densities are $5\times10^{11}$~cm$^{-2}$ (solid green line) and $2\times10^{12}$~cm$^{-2}$ (dashed-dotted black line). The dotted cyan line shows the excitonic DOS function without resident electrons, where $\varepsilon_{\text{1s}}$ is the zero reference energy. The oscillatory behavior of the excitonic DOS function at positive energies is a result of the excitonic band structure when the electrons form a Wigner square lattice \cite{VanTuan_PRB23L}. This oscillatory behavior is suppressed when the resident carriers are randomly distributed, as shown by the dashed magenta line in Fig.~\ref{fig:dos}(b) for $2\times10^{12}$~cm$^{-2}$. The trion energy is broadened and its resonance amplitude weakens when the resident carriers are randomly distributed, but its resonance energy does not shift because the trion size is smaller than the average distance between resident carriers at these densities. 

The solid black line in Fig.~\ref{fig:dos}(c) shows the calculated energy blueshift of the exciton resonance as a function of the density of distinguishable resident electrons. To compare this blueshift with the case of localized indistinguishable resident electrons, the sign of the interaction between the electron and point-like dipole is switched \cite{VanTuan_PRB23}, such that the dipole-charge potential is repulsive. The  solid and dashed red lines in Fig.~\ref{fig:dos}(c) show the energy blueshift of the exciton resonance as a function of the density of indistinguishable resident electrons when they form a Wigner square lattice or randomly distributed in the monolayer, respectively. Similar to the experimental result in Fig.~\ref{fig:1s}, distinguishable resident electrons induce stronger energy blueshift, stemming from the presence of an optimal trion state at a lower energy. 

}

\section{Compressibility and exciton resonances} \label{sec:point2}
In previous sections, we have implicitly assumed that the two-dimensional gas of mobile carriers is compressible, meaning that the resident carriers can scatter in response to absorption or emission of the e-h pair (e.g., to accommodate the Pauli exclusion principle). This response is impeded in TMD heterostructures with fractionally filled moir\'{e} lattices or in semiconductors with LLs at integer fillings wherein the macroscopic state of the resident carriers is incompressible. That is, when the system enters an insulating ground state with a finite energy gap below the excited state. In the following, we first present relevant experimental results and then discuss how compressibility of the resident carrier gas is related to the amplitude and energy shift of excitonic resonances.

\begin{figure*}
\includegraphics[width=15.5cm]{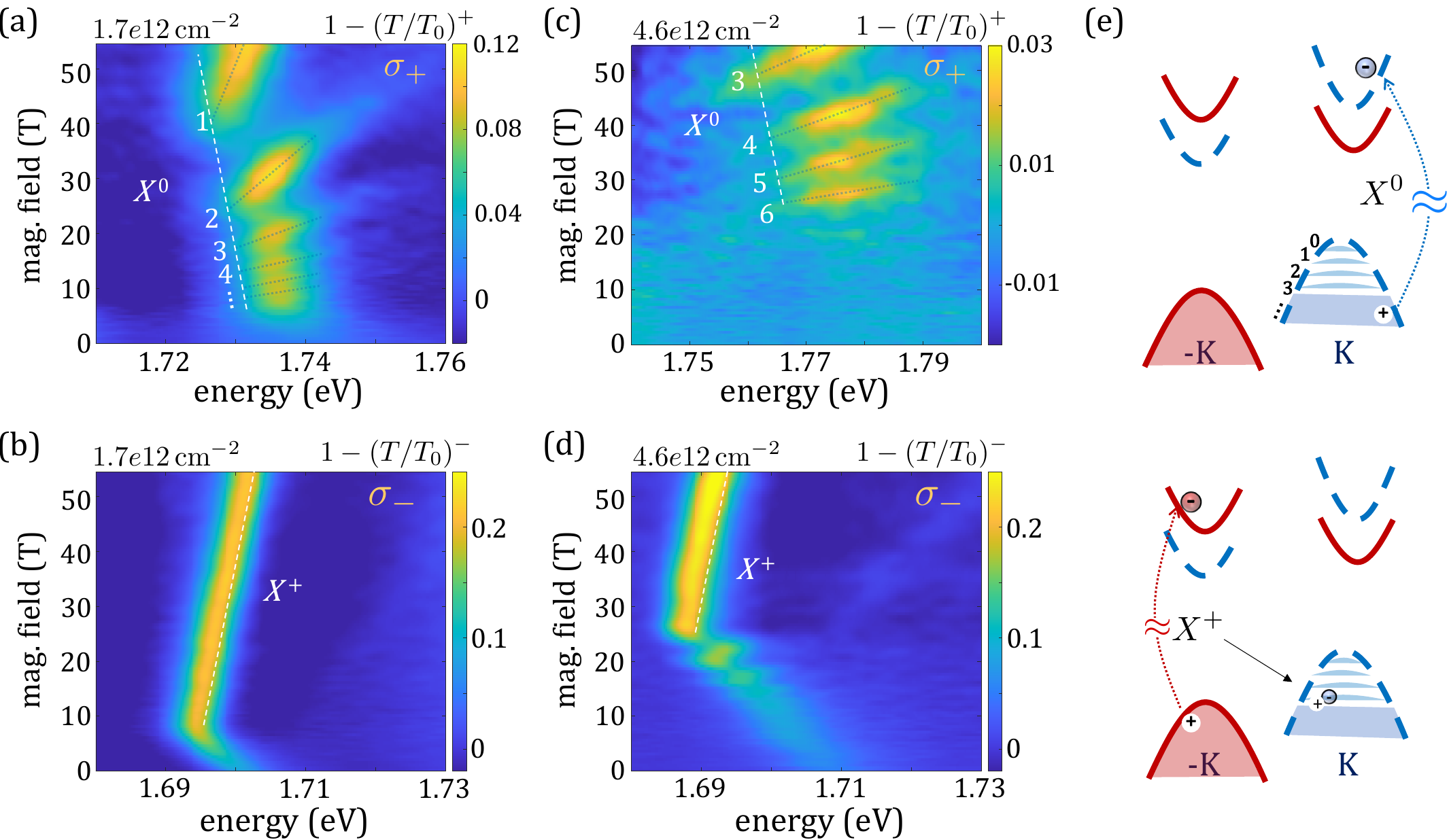}
 \caption{Colormaps of helicity resolved magneto-optical absorption spectra of an electrostatically hole-doped WSe$_2$ monolayer as a function of magnetic field and photon energy at 4$\,$K. Taken from Ref.~\cite{Li_NanoLett22}. The measured signal is $(1-T/T_0)^\pm$,  where $T$ ($T_0$) is the transmission (reference) spectrum, and $\pm$ refers to light with $\sigma_\pm$ polarization. The hole density is 1.7$\times10^{12}$~cm$^{-2}$ in (a)-(b) and 4.6$\times10^{12}$~cm$^{-2}$ in (c)-(d). (e) Schemes of optical excitations in valley-polarized WSe$_2$ monolayer with hole Landau levels at the VB valley of $K$. }\label{fig:T}
\end{figure*}

Starting with a monolayer system, Fig.~\ref{fig:T} shows $B$-dependent polarized optical absorption spectra of hBN-encapsulated WSe$_2$ \cite{Li_NanoLett22} (we note that these colormaps show $B$-dependent spectra at fixed carrier density, in contrast to density-dependent spectra at fixed $B$ shown in previous figures). The hole density in Figs.~\ref{fig:T}(a) and (b) is 1.7$\times10^{12}$~cm$^{-2}$. The emergence of exciton resonances in Fig.~\ref{fig:T}(a) and the onset of energy blueshift of $X^+$ in Fig.~\ref{fig:T}(b) take place when $B\gtrsim10$~T. The hole density in Figs.~\ref{fig:T}(c) and (d) is 4.6$\times10^{12}$~cm$^{-2}$, for which a similar behavior is observed when $B\gtrsim25$~T.  These strong magnetic fields are needed to fully valley-polarize the monolayer where holes only populate LLs of the VB valley at $K$, as shown in Fig.~\ref{fig:T}(e). Photoexcitation of this valley creates optimal excitons with indistinct resonance ($\sigma_+ \rightarrow X^0$), while photoexcitation of the valley at $-K$ creates optimal positive trions with distinct resonance ($\sigma_- \rightarrow X^+$).

The energy blueshift of $X^+$ when $B$ increases in Figs.~\ref{fig:T}(b) and (d) can be understood from the larger $g$-factors of the VB  \cite{Robert_PRL21}, where the Zeeman energy lowers the VB valley at $-K$ more than it lowers the CB valleys at $-K$ (see Fig.~\ref{fig:T}(e)). The dashed white lines in Figs.~\ref{fig:T}(a) and (c) trace the energy redshift from the optical transition in $K$. They have opposite slope to the energy blueshift in Figs.~\ref{fig:T}(b) and (d) due to the opposite-sign $g$ factors of the opposite valleys, $g_{-K} = -g_{K}$. That is, the Zeeman energy raises the VB valley at $K$ more than it raises the CB valleys at $K$. The dashed white lines in this case trace the energy redshift of $X^0$ at exact integer filling of the LLs, where the labeled integers indicate the highest filled LL of the VB valley at $K$. The indistinct resonances of the optimal exciton in Fig.~\ref{fig:T}(c) appear above 1.76~eV whereas those in Fig.~\ref{fig:T}(a) appear below 1.74~eV due to different numbers of filled LLs  (i.e., weaker/stronger role of Pauli blocking at smaller/larger hole densities).

The unique behavior we wish to focus on is the repeated blueshift and amplitude modulation of the resonance $X^0$ between exact integer filling of the LLs, shown when the exciton resonance departs from the white dashed line and extends into the dotted darker lines in Figs.~\ref{fig:T}(a) and (c). We glean two important insights from these measurements. 

\begin{enumerate}

\item The absorption amplitude of $X^0$ nearly vanishes at integer LL fillings and reaches a maximum at half fillings. This behavior can be seen by comparing the amplitudes along the dashed white line and those midway through the dotted black lines in Figs.~\ref{fig:T}(a) and (c). In contrast,  Figs.~\ref{fig:T}(b) and (d)  show that the absorption amplitude of $X^+$ is hardly affected by the filling factor, consistent with the fact that the photoexcited pair is distinguishable and holes are fully polarized. That is, because $X^+$ is a distinct resonance of an optimal positive trion. 

\item When the $(\ell+1)$ LL is partially filled, the energy blueshift of the exciton resonance along a black dotted line is qualitatively described by 
\begin{equation}
E_{X^0}(B) \approx E_{\ell+1 } +  C \times  \frac{B - B_{\ell+1}}{B_{\ell}-B_{\ell+1}}\,,  \label{eq:Eb}
\end{equation}
where $B_\ell \geq B \geq B_{\ell+1}$. $B_\ell$ and $B_{\ell+1}$ are the magnetic fields needed to  fill exactly $\ell$ and $(\ell+1)$ LLs, respectively. $E_{\ell+1 }$ is the exciton resonance energy when exactly ($\ell+1$) LLs are filled, and  $C$ is an energy constant that depends on charge density. 

\end{enumerate}

\begin{figure*}
\includegraphics[width=15.5cm]{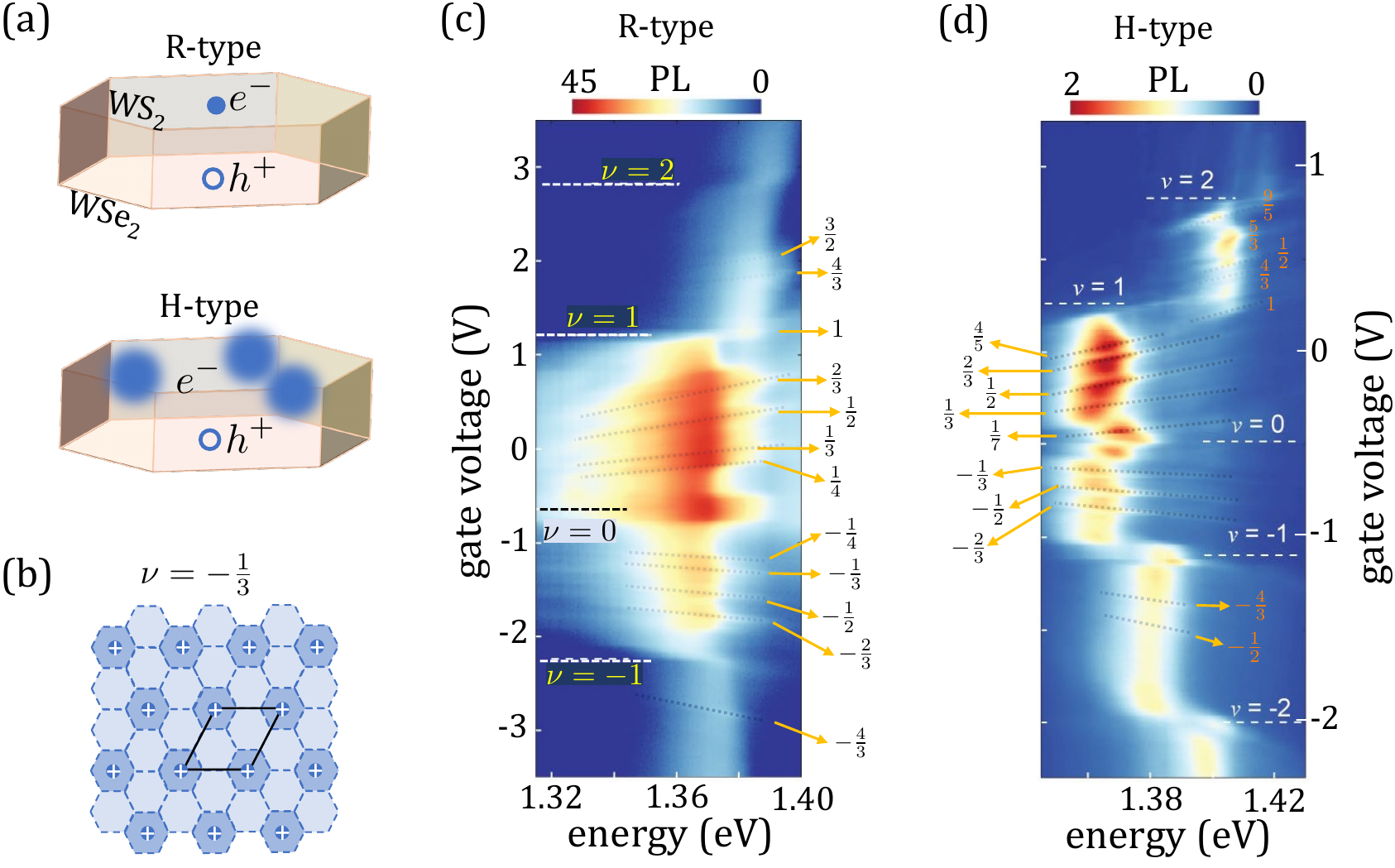}
 \caption{(a) Schemes of interlayer exciton complexes within the moir\'{e} unit cell of WS$_2$/WSe$_2$ heterobilayers. (b) Scheme of the moir\'{e} superlattice when the fractional hole filling is $\nu=-1/3$. (c) and (d) Low temperature PL intensity plots of the interlayer exciton versus gate voltages in R- and H-stacked devices, respectively. Taken from Ref.~\cite{Wang_NatMater23}. The corresponding optical excitations are 1.678 and 1.96~eV, and the laser powers are 50 nW. These measurements were taken at zero magnetic field. }\label{fig:moire}
\end{figure*}

A behavior with similar trends is observed in PL experiments of WS$_2$/WSe$_2$ heterobilayers performed by Wang \textit{et al.}  \cite{Wang_NatMater23}. Figure~\ref{fig:moire}(a) shows schemes of the interlayer exciton complex in the moir\'{e} unit cell.  The top scheme corresponds to R-type stacking in which the electron and hole are aligned. The bottommost CB of the heterobilayer belongs to the top monolayer (WS$_2$),  and the topmost VB belongs to the bottom monolayer (WSe$_2$).  The bottom scheme corresponds to H-stacking (60$^\circ$ twist angle between WS$_2$ and WSe$_2$), wherein the lowermost part of the CB  in the top monolayer is spread across the moir\'{e} unit cell \cite{Wang_NatMater23}. Figure~\ref{fig:moire}(b) shows a scheme of the moir\'{e} superlattice when the fractional hole filling is $\nu=-1/3$. The highlighted parallelogram encompasses the area of three moir\'{e} unit cells, within which the total charge is one hole. 

The rich PL behavior of the interlayer moir\'{e} excitons is shown in Figs.~\ref{fig:moire}(c) and (d) as a function of gate voltage and photon energy for R- and H-stacked devices, respectively \cite{Wang_NatMater23}. The horizontal white dashed lines correspond to gate voltages with integer filling of the moir\'{e} unit cells. The gate voltage at charge neutrality is marked by $\nu=0$. At larger (smaller) gate voltages, electrons (holes) are added to the WS$_2$ (WSe$_2$) monolayer and $\nu>0$ ($\nu<0$). The black dotted lines in Figs.~\ref{fig:moire}(c) and (d)  trace the energy blueshift of the interlayer moir\'{e} exciton when electrons or holes are added to the monolayer. The left edges of these black dotted lines mark voltages of exact fractional filling in the moir\'{e} superlattice.  The PL behavior of the interlayer exciton in Fig.~\ref{fig:moire} when changing the charge density is not foreign to the absorption behavior of $X^0$ in Fig.~\ref{fig:T}  when changing the magnetic field. Namely, 

\begin{enumerate}

\item The PL intensity of the interlayer moir\'{e} exciton nearly vanishes at exact fractional filling, and reaches maximum intensity when the charge densities are between two fractional states  (mid section of the black dotted lines). 

\item The energy blueshift of the interlayer moir\'{e} exciton along the black dotted lines in Figs.~\ref{fig:moire}(c) and (d) can be described as 
\begin{equation}
E_{IX}(n) \approx E_{\nu'} +  \mathcal{C} \times  \frac{n - n_{\nu'}}{n_{\nu}-n_{\nu'}}\,,  \label{eq:En}
\end{equation}
where $n_\nu \geq n \geq n_{\nu'}$. $n$ is the charge density, limited between the charge densities of adjacent fractionally-filled states, $n_\nu$ and $n_{\nu'}$ ($|\nu| > |\nu'|$). Since charge density is a positive quantity, we use assignments such as $\{\nu =1/3,\,\nu'=1/4\}$ in the electron-doped case versus $\{\nu =-1/3,\,\nu'=-1/4\}$ in the hole-doped case. $E_{\nu'}$ is the resonance energy at fractional filling $\nu'$ (left edges of the black dotted lines), and $\mathcal{C}$ is an energy constant that depends on the charging energy of the moir\'{e} unit cell. 

\end{enumerate}

\subsection*{Suggested interpretation}
The experimental findings show that the absorption amplitude of the neutral exciton vanishes at exact integer filling of LLs in a TMD monolayer, and that the emission amplitude of the interlayer exciton vanishes at exact fractional filling of the TMD moir\'{e} heterobilayer.  Both cases suggest that the exciton cannot exist if the state of the resident carriers is incompressible. To accommodate the presence of the exciton in a monolayer semiconductor, indistinguishable resident carriers must be able to elastically scatter out of the area occupied by the exciton or else the Pauli exclusion principle is violated \cite{Liu_PRL20,VanTuan_PRB23}. We emphasize that the presence of resident carriers is not necessary for emission or absorption of an exciton; these processes are strongest at charge neutrality without any resident carriers. The argument is that if resident carriers are present in the system, then they must reside in a compressible state for an exciton to exist. 

The elastic scattering in the WS$_2$/WSe$_2$ heterobilayer system is needed because the repulsion between overlapping distinguishable electrons in the WS$_2$ monolayer is stronger than their attraction to a hole in the WSe$_2$ monolayer. In other words, the interlayer exciton is formed when one of the two electrons scatters out of the exciton region. Such scattering is feasible with compressible states (e.g., a non-fractional state in the moir\'{e} heterobilayer), where the maximum PL intensity occurs midway between two fractional states because elastic scattering is strongest at these conditions. That is, when $f(1-f)=1/4$ where $f=1/2$ and $(1-f)=1/2$ are the occupation probability of a resident carrier prior and after scattering, respectively. The resident carriers can then scatter relatively freely and enable the shakeup processes.

Similarly, exciton formation is inhibited in a monolayer subjected to LL quantization at integer filling. Without loss of generality, we explain this physics in a hole-doped monolayer (e.g., Figs.~\ref{fig:T}(a) and (c)). The hole of the exciton is not subjected to LL quantized motion because the exciton is  charge neutral \cite{Yasui_PRB95,VanTuan_arXiv25,Whittaker_PRB97}. At integer filling conditions, the holes in the LLs cannot make room for the indistinguishable hole of the exciton (with similar spin and valley). Having a partially filled LL alleviates this constraint because holes from this LL can hop between cyclotron orbit centers. That is, indistinguishable resident holes avoid the exciton by elastic scattering away from the photoexcited hole.  

Finally, we notice that the energy blueshift of $X^0$ along the dotted black lines in Figs.~\ref{fig:T}(a), ~\ref{fig:T}(c), ~\ref{fig:moire}(c) and  ~\ref{fig:moire}(d) are all related to the density of compressible resident carriers. This dependence is implicit in the case of a partially filled LL, denoted by the fraction term in Eq.~(\ref{eq:Eb}), where the density of resident carriers in a LL is proportionate to the magnetic field. This dependence in explicit in the moir\'{e} heterobilayer system, evidenced from the fraction term in Eq.~(\ref{eq:En}), which denotes the relative density of  itinerant resident carriers between adjacent incompressible states. Deriving the direct relation between compressibility and energy shift, through which one can better quantify the phenomenological $C$ terms in Eqs.~(\ref{eq:Eb})-(\ref{eq:En}), will be studied elsewhere. 


\section{conclusion}\label{sec:outlook} 

The systematic analysis presented in this work shows that distinguishability and optimality, particularly as they relate to the Pauli exclusion principle, are key factors in determining the energy shift and broadening of various excitonic resonances in electrostatically-doped semiconductor systems.  We have corrected common misconceptions, showing that screening by charge particles and band filling are in fact minor reasons for the broadening or energy blueshift of excitonic resonances.  The one manifestation of screening is the energy redshift of distinct resonances of optimal excitonic complexes, where Pauli exclusion plays minimal role. The screening leads to a reduction of the bandgap energy and a reduction of the binding energy, where the former is slightly larger than the latter, leading to a small net energy redshift when the charge density increases. This effect is seen through distinct resonances of optimal complexes since their excitation (or recombination) does not involve a shakeup of the resident carrier distribution, and thus, these resonances neither broaden nor decay. 

Shakeup processes do, however, govern the energy shifts and broadening of indistinct resonances of optimal excitonic complexes. In absorption, these processes break the spatial overlap between distinguishable and indistinguishable resident carriers, allowing the distinguishable carrier to bind to a photoexcited e-h pair. In emission, the shakeup process reunites distinguishable and indistinguishable resident carriers after the recombination. 

We have self-consistently analyzed the various trends of energy shifts and broadenings of neutral-exciton absorption resonances. {\color{black}The stronger energy blueshift of excitons in the presence of distinguishable resident carriers is caused by the transfer of density-of-states from the suboptimal complex (exciton) to the optimal one (trion). The enhanced broadening in this case is attributed to the ultrafast time it takes a suboptimal exciton to bind to a distinguishable resident carrier. The neutral exciton shows resilience in the absence of distinguishable resident carriers (i.e., smaller energy shift, weaker broadening, and stronger amplitude), rendering the exciton an optimal complex, and especially in moir\'{e} heterobilayer systems wherein its resonance can also become distinct.} Finally, we have elucidated the intimate connection of the compressibility of the resident carrier gas to the amplitude and energy shift of the neutral exciton resonance.

Quantifying and elucidating the details of shakeup processes are interesting directions for future investigations, testing whether collective spin excitations such as magnons in valley-polarized systems or collective charge excitations such as long-wavelength plasmons can take part in these processes. Similarly, further studies are needed to quantify the compressibility of the resident carriers in moir\'{e} heterobilayers from the measured energy shift and amplitude of exciton resonances. Hopefully, the pervading nature of the Pauli exclusion principle along with the ensuing concepts  of optimality of excitonic complexes, distinguishability of the e-h pair, distinctiveness of the optical resonances, and compressibility of the resident carrier gas will become the accepted approach to interpret optical studies in electrostatically-doped semiconductors. 


\acknowledgments{This work was mainly supported by the U.S. Department of Energy, Basic Energy Sciences, Division of Materials Sciences and Engineering under Award No. DE-SC0014349. Scott Crooker acknowledges support from the U.S. Department of Energy ``Science of 100~T'' program. The NHMFL is supported by National Science Foundation Grant No. DMR-1644779, the state of Florida, and the U.S. Department of Energy. Cedric Robert and Xavier Marie are supported by the Agence Nationale de la Recherche under the program ESR/EquipEx+ (Grant No. ANR-21-ESRE- 0025) and ANR projects ATOEMS and IXTASE.}

\appendix

{\color{black}
\section{Caveats}\label{sec:caveats}

Quantum numbers that determine the distinguishability of an electron or hole in a semiconductor are its energy band, spin and wavevector ($n,s,\mathbf{k}$). To that one should add the monolayer index in case of heterobilayers. When dealing with excitonic complexes, we use the valley index of the electron (or hole) instead of its wavevector. The reason is that a particle can bind to a complex by moving with respect to other particles in the complex, and this internal motion requires a finite bandwidth of wavevectors (low energy states) in the \textit{valley} at which the particle resides.

In general, the particles of an excitonic complex have to be mutually distinguishable, and they include resident electrons (holes) at the edge of the CB (VB) with different valley-spin configurations. Yet, there are exceptions to this rule. Biexcitons with two indistinguishable particles and two distinguishable ones can be stabilized only if the effective mass of the indistinguishable particles is at least five times larger than the effective mass of the distinguishable particles \cite{Mostaani_PRB17}. The relative motion of the lighter-mass distinguishable particles is much faster than that of the spatially non-overlapping indistinguishable particles, thereby suppressing the exchange effects between the heavy particles and keeping such biexciton bound.

The case of a bound trion that has two electrons (or holes) with the same spin and valley is more subtle. At zero or weak magnetic fields, such trion state can only emerge if the trion has  nonzero angular momentum \cite{Tiene_PRB22}. On the other hand, the optically active trions have zero angular momentum ($s$-states), and they cannot be stabilized with two particles that share the same spin and valley regardless of how heavy their mass may be. Such particles cannot be in contact with the opposite-charge particle simultaneously because of the Pauli exclusion principle, rendering the formation of a nodeless $s$-state impossible \cite{VanTuan_PRB25}.  An exception to this rule can happen by applying a strong magnetic field, such that the separation between LLs (cyclotron energy) exceeds the trion binding energy. This condition can be met in semiconductor quantum wells, where the binding energy of the trion is less than 5~meV \cite{Kheng_PRL93,Finkelstein_PRL95,Astakhov_PRB00,Bracker_PRB05,Huard_PRL00,Shields_PRB95,Finkelstein_PRL96,Homburg_JCG00, Finkelstein_PRB96, Whittaker_PRB97,Andronikov_PRB05}. A trion state can then be stabilized even if it includes two particles with the same band-spin-valley configuration, because the LLs behave as additional distinguishability knob \cite{Finkelstein_PRB96}. That is, the Pauli principle no longer provides a strong constraint as at zero field when the degeneracy of the LL is large enough \cite{Whittaker_PRB97}.}

\section{Hexcitons and oxcitons}\label{sec:hex}
Much of the way we perceive excitons in semiconductors is attributed to the seminal work of Elliott, who tailored the hydrogen model to calculate exciton states  \cite{Elliott_PR57}. Employing the semiconductor's dielectric constant and effective masses of its CB electron and VB hole, the two-body model successfully predicts the energy spectrum of charge-neutral excitons when using appropriate Coulomb potential models \cite{Cudazzo_PRB11,VanTuan_PRB18,Meckbach_PRB18}. Similar analogy can be drawn between a trion and hydrogen ion, H$^{-}$, where the trion complex is modeled as a three-body problem \cite{Stebe_PRL89,Esser_pssb01}. Solutions of this problem yield binding energies which agree with the measured energy difference between the trion and exciton resonances in the spectra of TMD monolayers \cite{Mayers_PRB15,Kylanpaa_PRB15,Kidd_PRB16,Donck_PRB17,Filikhin_Nano18,VanTuan_PRL19}. 

If we wish to generalize this idea further, the following subtlety should be borne in mind when dealing with larger excitonic complexes. The hydrogen ion $H^{-N}$, where $N$ is the number of extra electrons, is unstable when $N>1$. One proton cannot hold more than two electrons, and therefore $H^{-N}$ dissociates to $H^{-}$ and $N-1$ electrons that end up far away from the ion and from each other. On the other hand, the stability of an excitonic complex is not a problem when dealing with an electrostatically-doped semiconductor that has $N$ Fermi seas. Here, the photoexcited pair acts as a charge-neutral hydrogen atom, and it binds to $N$ electrons with different spin-valley configurations which are near each other before excitation. To explain this point, let us consider a semiconductor whose CB has $N=2$ populated valleys and the photoexcited pair is distinguishable, as shown in Fig.~\ref{fig:WSe2}(d). The average distance between nearby electrons, one from the valley at $K$ and the other from $-K$, is about $k_F^{-1}$ where $k_F$ is the Fermi wavenumber. When the electron density is large enough, the hexciton formation means that the photoexcited pair binds to these electrons which are already near each other at equilibrium. The same principle holds true when $N=3$ (oxciton) \cite{VanTuan_PRL22} or in multivalley semiconductors with $N \geq 4$.
%

\section{LL quantization of hexciton and oxciton resonances in a strong magnetic field}\label{sec:hob}

\begin{figure}[t!]
\centering
\includegraphics[width=8.5cm]{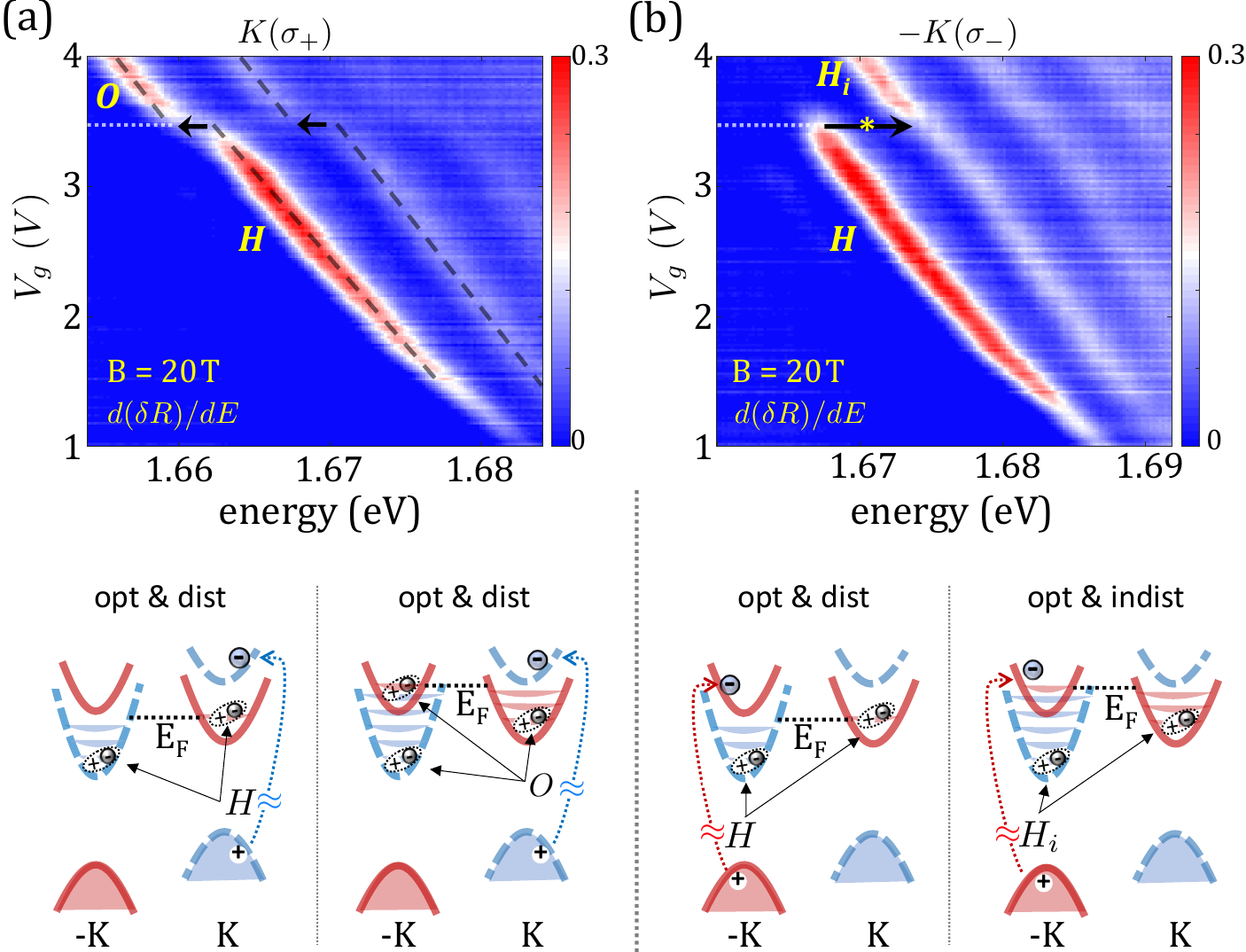}
\caption{Hexciton and oxciton resonances in electron-doped WSe$_2$ when the applied magnetic field is 20~T. Taken from Ref.~\cite{Wang_PRX20}. The bottom diagrams show corresponding absorption processes for each of the helicity-resolved optical transitions. }\label{fig:hob} 
\end{figure}

Figure~\ref{fig:hob} highlights the spectral window that pertains to the hexciton ($H$) and oxciton ($O$) resonances in Fig.~\ref{fig:dr}(c). Figure~\ref{fig:hob}(a) shows that the crossover from distinct hexciton to oxciton resonances at $V_g \gtrsim 3.5$ is corroborated by the reflectance signal with helicity $\sigma_-$ in Fig.~\ref{fig:hob}(b), which shows that its hexciton signal jumps to the next LL at higher energy \cite{Li_NanoLett22}. The left-pointing black arrows in Fig.~\ref{fig:hob}(a) reflect the increased binding energy of the oxciton complex compared with that of the hexciton. On the other hand, the right-pointing black arrow with a yellow asterisk in Fig.~\ref{fig:hob}(b) signifies the onset of Pauli blocking where the hexciton resonance becomes indistinct (the top valley of $-K$ starts to be filled). 

Unlike the absorption resonances of exciton and trion complexes in Figs.~\ref{fig:dr}(c) and (d), the resonances $H$ and $O$ have multiple LL branches at a given electron density, as highlighted in Fig.~\ref{fig:hob}. This unique behavior was explained in Ref.~\cite{VanTuan_PRL22} by considering the structure of the hexciton (or oxciton), which consists of a dark trion in its core and satellite electron(s) from the optically-active top valleys. The VB hole strongly binds to the two electrons from the bottom valleys because they are $\sim$30\% heavier than electrons of the top valleys \cite{Kormanyos_2DMater15}. The implication is  that the fast relative motion of an electron in a tightly bound trion cannot be quantized in LLs like that of a free electron. On the other hand, the relative motion of a satellite electron is slower, and as such, it has larger resemblance to the LL quantized motion of a free electron. 

This behavior supports the observed LL branches in Fig.~\ref{fig:hob}(b) and the observation that the {\color{black} indistinct} resonance $H_i$ continues to redshift despite being indistinct (Sec.~\ref{sec:shakeup}). Namely, the photoexcitation does not require a shakeup process to accommodate the Pauli exclusion principle if the optically-active electron is promoted to an empty LL level and its quantized motion resembles that of a free electron.

{\color{black}
\section{Energy shifts of optical gaps}\label{sec:optgap}

The energy shift of a resonance, which one measures in experiment, reflects a change in the optical gap of the e-h pair that undergoes optical transition. When an excitonic complex is formed, it inherits the renormalized self energies of the resident carriers to which the e-h pair binds. Since these renormalized energies are essentially the same before and after the absorption, they cannot be measured through the absorbed photon energy. 

To support this understanding, we focus on the Zeeman energy shift of the distinct resonance $X^+$ of the optimal positive trion in Figs.~\ref{fig:T}(b) and (d). This energy shift can be evaluated very well from $(g_v - g_c)\mu_BB$, where $g_c = 3.84$ and $g_v = 6.1$ are the single-particle $g$ factors of the valleys that take part in the optical transition \cite{Robert_PRL21}. The exchange-induced renormalization energy of the hole to which the pair binds is not probed by the photon energy, as one can realize by the fact that the same slope is measured in Figs.~\ref{fig:T}(b) and (d), albeit the hole density is ramped up from $1.7\times 10^{12}$ in Fig.~\ref{fig:T}(b) to $4.6\times 10^{12}$ in Fig.~\ref{fig:T}(d). That is, the measured photon energy only reflects the rigid Zeeman shifts at the edges of the CB and VB valleys that take part in the optical transition. 

This understanding can also be sharpened by contrasting this result with studies that report `enhanced' $g$-factor values when the resident carriers are fully polarized \cite{Liu_PRL20,Oreszczuk_2DMater13,Krishtopenko_JPCM11}. For example, the `enhanced' $g$-factor of holes in a fully valley-polarized WSe$_2$ monolayer can be realized from the behavior of $X^+$ in Fig.~\ref{fig:dr}(c). At $B=20\,$T, the holes in this device remain fully valley-polarized up to a density of $n_h=2.5\times 10^{12}$~cm$^{-2}$ (corresponding to $V_g = -1.7$~V). This result means that the Zeeman energy difference between the $\pm K$ valleys, $E_Z = 2g_v\mu_BB$, exceeds the Fermi energy $E_F = 2\pi \hbar^2 n_h /m_h$, where the holes populate one valley and their effective mass is $m_h =0.36m_0$ \cite{Kormanyos_2DMater15}. While the condition $E_Z \gtrsim E_F$ yields  $g_v \gtrsim 14$, one measures $g_v \simeq 6$ from the energy shift of a trion resonance as a function of magnetic field \cite{Robert_PRL21}. The latter reflects a single-particle  $g$-factor that one can also extract from first-principle calculations \cite{Forste_NatComm20}. The conclusion is that the exchange interaction, which is responsible for the `enhanced' $g$-factor through the self-energy of a resident hole  \cite{GiulianiVignale_Book}, is not observed through the optical transition of an excitonic complex. This supports the understanding that the renormalized self-energies of resident carriers to which the e-h pair binds are essentially the same before and after the absorption, and therefore, they cannot be measured through the absorbed photon energy.}

\section{Alternative explanation to the universal energy redshift}\label{sec:alternative}

In  Refs.~\cite{VanTuan_PRL22,VanTuan_PRB22} , we have attributed the sustained energy redshift of hexcitons when the charge density increases to the compression of Coulomb holes that surround the hexciton. We did not pursue this explanation in the main text of this work because it raises questions when we examine Fig.~\ref{fig:dr}(c). Explicitly, the energy redshift of the resonance $O$ is similar to that of the resonance $H$, and both shifts are similar in magnitude to the redshift of the resonance $X^+$ (but with opposite slope). Since the energy redshifts of $X^+$, $H$ and $O$ have similar magnitudes while their underlying excitonic complexes include one, two and three Coulomb holes, respectively, the compression of Coulomb holes is not a convincing explanation.

{\color{black}\section{Trion-plasmon interaction during absorption and emission}\label{sec:trion_plasmon}
The plasmon-assisted trion absorption (or emission) is a process wherein the optical transition is accompanied by the generation of a plasmon wave. As discussed in Sec.~\ref{sec:shakeup}, generation of a plasmon wave is one way to facilitate the breakup (reunion) shakeup process during absorption (emission) of an indistinguishable e-h pair in an optimal complex. Without loss of generality, we analyze the plasmon-trion absorption case in electron-doped monolayer. The modifications needed to describe the hole-doped case are straightforward, and the modifications needed to describe the emission case will be discussed later. 

\begin{widetext}
We employ second-order perturbation theory to quantify the rate of the plasmon-assisted trion absorption process,
\begin{eqnarray}
\tau_{\mathbf{k},\Omega}^{-1} &=& \frac{2\pi}{\hbar} \sum_{\mathbf{q}} \left| \sum_{\mathbf{k}'} \frac{ \langle e_{\mathbf{k}}, \Omega  | H_{\text{LM}}| T_{\mathbf{k}'}, 0\rangle  \langle T_{\mathbf{k}'},0 | H_{\text{e-pl}} | T_{\mathbf{k}-\mathbf{q}},\omega_\mathbf{q} \rangle}{\hbar \omega_{\mathbf{q}} + \varepsilon_{T, \mathbf{k}-\mathbf{q}} - \varepsilon_{e,{\mathbf{k}}} } \right|^2 \delta\left( \hbar \Omega + \varepsilon_{e,{\mathbf{k}}} - \varepsilon_{T,{\mathbf{k}-\mathbf{q}}} - \hbar \omega_{\mathbf{q}} \right) \nonumber \\
\, &=& \frac{A}{2\pi\hbar} \int \! d^2 \mathbf{q} \left| \frac{M_{\text{LM,T}}(\mathbf{k}) M_{\text{e-pl}}(\mathbf{k},\mathbf{q})} {\hbar \omega_{\mathbf{q}} + \varepsilon_{T,{\mathbf{k}-\mathbf{q}}} - \varepsilon_{e,{\mathbf{k}}} + i\Gamma } \right|^2  \, \frac{\Gamma/\pi}{\left( \hbar \Omega + \varepsilon_{e,{\mathbf{k}}} - \varepsilon_{T,{\mathbf{k}-\mathbf{q}}} - \hbar \omega_{\mathbf{q}} \right)^2 + \Gamma^2}, \label{eq:tp2nd}
\end{eqnarray}
\end{widetext}
where $A$ is the sample area. The initial state, $|  e_{\mathbf{k}}, \Omega \rangle $, comprises an electron with wavevector $\mathbf{k}$ and a photon, whose respective energies are  $\varepsilon_{e,{\mathbf{k}}}$ and $\hbar\Omega$. The final state, $|  T_{\mathbf{k}-\mathbf{q}}, \omega_{\mathbf{q}} \rangle $, comprises a trion with translational wavevector $\mathbf{k}-\mathbf{q}$ and emitted plasmon with wavevector $\mathbf{q}$, whose respective energies are  $\varepsilon_{T,{\mathbf{k}-\mathbf{q}}}$ and $\hbar\omega_{\mathbf{q}}$. The intermediate state, $|  T_{\mathbf{k}'}, 0 \rangle $,  comprises the created trion and zero plasmons, where we have assumed that the trion picks up the crystal momentum of the electron to which it binds because of the negligible momentum carried by the photon (i.e., $\mathbf{k}' = \mathbf{k}$). In the transition from the first to second lines of Eq.~(\ref{eq:tp2nd}), we have replaced the delta function with a Lorentzian of width $\Gamma$, and a similar broadening is applied to the energy denominator. $M_{\text{LM,T}}(\mathbf{k})$ is the light-matter matrix element of the trion optical transition, and $M_{\text{e-pl}}(\mathbf{k},\mathbf{q})$ is the matrix element of the plasmon-trion interaction \cite{Caruso_PRB16}
\begin{eqnarray}
M_{\text{e-pl}}(\mathbf{k},\mathbf{q}) &=& \sqrt{  \frac{\kappa_{\mathbf{q}}V_{\mathbf{q}} \hbar \omega_{\mathbf{q}}}{2\epsilon_{\mathbf{q}}[q\epsilon_{\mathbf{q}} +  \kappa_{\mathbf{q}}]}}. \label{eq:kappa_rpa}
\end{eqnarray}
$\kappa_q$ is the screening wavenumber due to the electrostatic doping, $V_{\mathbf{q}} = 2\pi e^2/A\epsilon_{\mathbf{q}} q$ is the Coulomb potential in two dimensions, and  $\epsilon_{\mathbf{q}}= \epsilon_d+r_0 q$ is the Keldysh-Rytova dielectric screening function \cite{Rytova_PM67,Keldysh_JETP79}, where $\epsilon_d$ is the effective dielectric constant of the environment, and $r_0$ is the polarizability of the monolayer \cite{Cudazzo_PRB11}. Using the random-phase approximation, the screening wavenumber and plasmon energy read \cite{Scharf_JPCM19},
\begin{eqnarray}
\kappa_{\mathbf{q}} &=& g_s g_\nu \frac{e^2 m}{\hbar^2 \epsilon_{\mathbf{q}}} \left[ 1 - \sqrt{1-\left(\frac{2k_F}{q}\right)^2} \theta(q-2k_F)\right],\nonumber \\
\omega_{\mathbf{q}} &=& \sqrt{\frac{q}{m \epsilon_{\mathbf{q}}} }ek_F. \label{eq:kappa}
\end{eqnarray}
$g_{s}=1$ and $g_\nu=2$ are the spin and valley  degeneracies in TMD monolayers, respectively, $m$ is the effective mass of the resident carriers, $\theta$ is the Heaviside step function, and $k_F=\sqrt{4\pi n/g_s g_\nu}$ is the Fermi wavenumber, where $n$ is the charge density.

To compare the plasmon-assisted trion absorption with the direct trion absorption (i.e., without plasmons), we write the latter rate as
\begin{eqnarray}
\widetilde{\tau}_{\mathbf{k},\Omega}^{-1} &=& \frac{2\pi}{\hbar} \left| M_{\text{LM,T}}(\mathbf{k}) \right|^2  \, \frac{\Gamma/\pi}{\left( \hbar \Omega + \varepsilon_{e,{\mathbf{k}}} - \varepsilon_{T,{\mathbf{k}}}  \right)^2 + \Gamma^2}\,. \,\,\,\,\,\,\,\,\,\,\,\,\,
\end{eqnarray}
The dotted lines in Figs.~\ref{fig:breakup}(e) and \ref{fig:reunion}(b) show the renormalized form of this expression at $k=0$ where the broadening parameter is $\Gamma=1$~meV.  Relative to this renormalized rate, the solid lines in Fig.~\ref{fig:breakup}(e) show the amplitude of the plasmon-assisted trion absorption from Eq.~(\ref{eq:tp2nd}) at $k=0$ for 4 different holes densities. The parameters of the hBN-encapsulated MoSe$_2$ monolayer are $r_0 = 4.1$~nm and $\epsilon_d = 3.8$ \cite{VanTuan_PRB18}, and the effective masses are $m_h=0.6m_0$ and $m_e=0.5m_0$ \cite{Kormanyos_2DMater15}, where the translational mass of the positive trion is $M=2m_h+m_e=1.7m_0$. In addition, we assume parabolic energy dispersions, where the zero energy reference level is that of the idle trion at zero charge density. Repeating the calculation for the emission spectrum in Fig.~\ref{fig:reunion}(b) is straightforward, where the second line of Eq.~(\ref{eq:tp2nd})  is evaluated with the following replacements, $\hbar\omega_{\mathbf{q}} \rightarrow -\hbar\omega_{\mathbf{q}} $, $\varepsilon_{T,{\mathbf{k}-\mathbf{q}}}  \rightarrow \varepsilon_{T,{\mathbf{k}}} $, and $\varepsilon_{e,{\mathbf{k}}}  \rightarrow \varepsilon_{e,\mathbf{k}-\mathbf{q}} $.

Finally, the random-phase approximation (RPA) tends to overestimate the role of screening. To mitigate this effect, one can replace the screening wavenumber  $\kappa_q$ in Eq.~(\ref{eq:kappa_rpa}) with $\eta_n \kappa_q$, where $\eta_n = \exp(-n/n_0)$ \cite{Liu_PRL20}. The limit $\eta_n=0$ at which $n_0 \rightarrow 0$ completely neglects the role of screening (i.e., no plasmons), whereas ideal RPA  is applicable in the limit $\eta_n=1$ at which $n_0 \rightarrow \infty$. The results in Figs.~\ref{fig:breakup}(e) and \ref{fig:reunion}(b) were evaluated with $n_0=5\times10^{11}$~cm$^{-2}$. If we were to use the ideal RPA case ($\eta_n=1$), the  plasmon-assisted processes in Figs.~\ref{fig:breakup}(e) and \ref{fig:reunion}(b) exceed the direct processes (i.e., the rates shown by the solid lines are stronger that those of the dotted lines). }



\color{black}
\section{Quasi-localization}\label{sec:localization}
There are several supporting observations that strengthen the hypothesis of quasi-localization of resident carriers at small charge densities in TMD monolayers. First, interfacial fluctuations are inevitable in every sample due to disorder and imperfections in the monolayer or its vicinity. Such fluctuations act as shallow potential wells that trap resident carriers \cite{Zrenner_PRL94,Martin_NatPhys08}. This effect fades away when (i) the charge density is large enough to screen out these traps \cite{Li_PRM21}, or (ii) when the cyclotron radius due to a strong magnetic field is smaller than the radius of a typical shallow trap (that is, when the cyclotron energy is larger that the binding energy to the trap). Second, the estimated diffusion constants of thermalized singlet and triplet trions in hBN-encapsulated WSe$_2$ monolayer are of the order of 1~cm$^{2}$/s \cite{Beret_PRB23}, whereas the diffusion constant of excitons is found to improve from 2~cm$^{2}$/s to 40~cm$^{2}$/s when traps are screened out \cite{Rieland_NanoLett24}. Third, if triplet and singlet trions were itinerant, they would readily form hexcitons by binding to distinguishable electrons from the other bottom valley.  Yet, we do not observe hexciton emission at small electron densities in WSe$_2$ (or WS$_2$) monolayers. Finally, quasi-localization of the resident carriers at small charge densities supports the observation that trions hardly shift in energy \cite{VanTuan_PRB23L,VanTuan_PRB23}, in accord with the behavior of $X^-_{S,T}$ in Fig.~\ref{fig:wse2case}.

The quasi-localization has another subtle signature on the emission spectra in Figs.~\ref{fig:breakup}(b) and ~\ref{fig:wse2case}(a). While the energy redshift of indistinct resonances of optimal complexes exists at all charge densities, the redshift and broadening are enhanced at elevated charge densities. The enhancement in the case of $X^\pm$ in Fig.~\ref{fig:breakup}(b) is evident when $|V_g| \gtrsim 4\,$$V$, and in the case of $D^-$ and $D^-_B$ in Fig.~\ref{fig:wse2case}(b) when $V_g \gtrsim 15\,$$V$. Our suggested explanation is that reunion processes are relevant regardless of charge density, where the extent of the energy redshift and broadening depend on whether the resident carriers are quasi-localized or itinerant. In more detail, since the Pauli exclusion principle precludes overlap of indistinguishable resident carriers with trions, the rate at which indistinguishable resident carriers creep into the trion region after recombination depends on their itinerancy.  The smaller rate of energy redshift and broadening when the resident carriers are quasi-localized is consistent with the fact that Coulomb correlations help indistinguishable and distinguishable resident carriers to avoid each other better at small charge densities \cite{VanTuan_PRB23}. The resulting overlap between indistinguishable and distinguishable resident carriers is smaller than the one expected if they were itinerant, leading to mitigated redshift and broadening effects.

\begin{figure*}[t!]
\centering
\includegraphics[width=16cm]{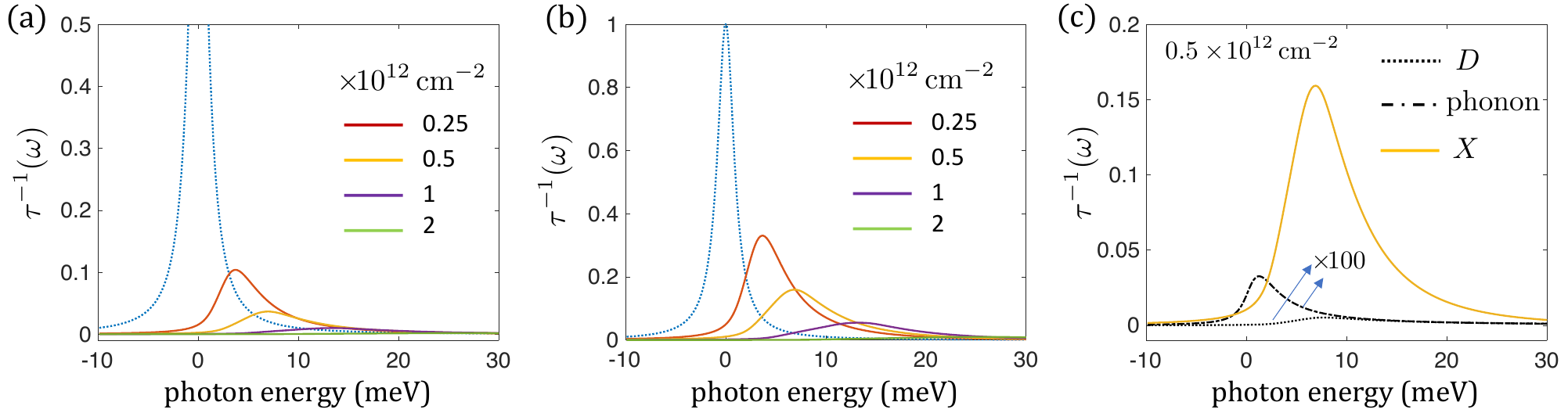}
\caption{{\color{black}(a) Calculated rates of exchange-assisted exciton absorption  in hole-doped WSe$_2$ monolayer (solid lines). The calculated rates are normalized with respect to the rate of the regular exciton absorption in the intrinsic limit (dotted line), where the photon energies are measured with respect to its resonance. (b) The same as in (a) but with using the unscreened Coulomb potential instead of the RPA screened Coulomb potential. (c) Calculated rates of exciton absorption in a WSe$_2$ monolayer, assisted by acoustic phonons (dashed-dotted line), direct Coulomb interaction of the exciton and holes (dotted line), and the exchange interaction of the exciton and holes (solid line). The former two are magnified by a factor of 100, and the latter two are calculated when the density of indistinguishable holes is $5\times10^{11}$~cm$^{-2}$.}}\label{fig:Xe} 
\end{figure*}

{\color{black}\section{Exchange-assisted exciton absorption}\label{sec:exc_x0}
Unlike the trion-plasmon interaction, the exciton-plasmon interaction is relatively weak because of the charge neutrality of the exciton. Instead, the exchange interaction between an exciton and a resident carrier becomes a dominant mechanism, where exchange interaction in this context means that the electron (hole) of the exciton is exchanged with a resident electron (hole) after their interaction \cite{Ramon_PRB03,Shahnazaryan_PRB17,Yang_PRB22}. 

We assume an electron-doped monolayer in the derivation below, where the switch to hole doping is straightforward. Similar to the procedure in Appendix~\ref{sec:trion_plasmon}, we employ second-order perturbation theory to quantify the rate of the exchange-assisted exciton absorption process, 
\begin{widetext}
\begin{eqnarray}
\tau_{\Omega,X}^{-1} &=& \frac{2\pi}{\hbar} \sum_{\mathbf{k},\mathbf{q}} \left| \sum_{\mathbf{p}} \frac{ \langle  \Omega, e_{\mathbf{k}}  | H_{\text{LM}}| X_{\mathbf{p}},e_{\mathbf{k}} \rangle  \langle X_{\mathbf{p}},e_\mathbf{k} | H_{\text{X-e}} | X_{\mathbf{p}+\mathbf{q}},e_{\mathbf{k}-\mathbf{q}} \rangle}{ \varepsilon_{X, \mathbf{p}+\mathbf{q}} + \varepsilon_{e,{\mathbf{k}-\mathbf{q}}} - \varepsilon_{X, \mathbf{p}} - \varepsilon_{e,{\mathbf{k}}}  } \right|^2 \!\!\!f_{\mathbf{k}}(1\!-\! f_{\mathbf{k}-\mathbf{q}})(1\!-\!f_{\mathbf{p}+\mathbf{q}}) \,\,\delta\left( \hbar \Omega \!+\! \varepsilon_{e,{\mathbf{k}}} \!-\!  \varepsilon_{X, \mathbf{p}+\mathbf{q}} \!-\! \varepsilon_{e,{\mathbf{k}-\mathbf{q}}}  \right) \nonumber \\
&=& \frac{2\Gamma}{\hbar} \sum_{\mathbf{k},\mathbf{q}} \left|  \frac{   M_{\text{LM,X}} \, M_{\text{X-e}}(0,\mathbf{k},\mathbf{q}) }{ \varepsilon_{X, \mathbf{q}} + \varepsilon_{e,{\mathbf{k}-\mathbf{q}}} - \varepsilon_{X, 0} - \varepsilon_{e,{\mathbf{k}}} + i\Gamma   } \right|^2 \frac{ f_{\mathbf{k}}(1- f_{\mathbf{k}-\mathbf{q}})(1-f_{\mathbf{q}})}{
\left( \hbar \Omega + \varepsilon_{e,{\mathbf{k}}} -  \varepsilon_{X, \mathbf{q}} - \varepsilon_{e,{\mathbf{k}-\mathbf{q}}} \right)^2 + \Gamma^2 }\,\, . \label{eq:xe2nd}
\end{eqnarray}
\end{widetext}
As before, we have assumed that the photon carries negligible momentum (i.e., $p=0$). The initial state is an electron with wavevector $\mathbf{k}$ and a photon with energy $\hbar\Omega$, and the final state is an exciton and electron with wavevectors $\mathbf{q}$ and $\mathbf{k}-\mathbf{q}$, respectively.  The Fermi distributions $f_{\mathbf{k}}(1- f_{\mathbf{k}-\mathbf{q}})$ ensure that the electron is scattered from a filled to an empty state. As discussed in Sec.~\ref{sec:blueX0indis}, the center-of-mass wavevector of the exciton in the final state does not overlap with that of an indistinguishable electron; hence the factor $(1-f_{\mathbf{q}})$.  $M_{\text{LM,X}}$ is the light-matter matrix element of the exciton optical transition, and $M_{\text{X-e}}(p\rightarrow 0,\mathbf{k},\mathbf{q})$ is the exciton-electron matrix element between an exciton in the light cone and an electron with wavevector $\mathbf{k}$. This matrix element has contributions from direct and exchange terms \cite{VanTuan_PRB25},
\begin{eqnarray}
M_{\text{X-e}}(0,\mathbf{k},\mathbf{q}) &=&  D({\bf q}) + X({\bf k,q}) \,\,,   \label{eq:MDX} \\
D({\bf q}) &=&  \frac{V_{{\bf q}}}{A} \sum_{{\bf q}'}  \phi_{\mathbf{q}' } \left[  \phi^*_{\bar{\eta}{\bf q}+{\bf q}'}    - \phi^*_{\eta{\bf q}+{\bf q}'} \right] \,, \nonumber \\
X({\bf k,q}) &=& \frac{1}{A} \sum_{{\bf q}'} V_{{\bf q}'} \phi_{\eta\mathbf{q} - \mathbf{k}- \mathbf{q}'} \left[ \phi^\ast_{\mathbf{q} - \mathbf{k}}  -  \phi^\ast_{\mathbf{q} - \mathbf{k}- \mathbf{q}'} \right] \,. \nonumber
 \end{eqnarray}
The Coulomb potential takes the form  $V_{\mathbf{q}} = 2\pi e^2/A\epsilon_{\mathbf{q}} q$ when the electrostatic screening is neglected, or $V_{\text{s},\mathbf{q}} = qV_{\mathbf{q}}/(q+\kappa_q)$ when the RPA screening is considered with the screening wavenumber $\kappa_q$ in Eq.~(\ref{eq:kappa}). As before, $\epsilon_{\mathbf{q}}= \epsilon_d+r_0 q$ is the Keldysh-Rytova dielectric screening function. $\phi_{k}$ is the Fourier transform of the exciton wavefunction, and we consider the 1$s$ exciton state of a 2D-hydrogen model, $\phi_{\mathbf{k}}= \sqrt{8\pi}a_x/[1 + (ka_x)^2]^{3/2}$ where $a_x$ is the exciton's Bohr radius. The other parameters are the mass ratios $\eta = m_e / (m_e + m_h)$ and $\bar{\eta}= \eta - 1$. 

To compare the exchange-assisted exciton absorption with the zero-line exciton absorption (i.e., without scattering), we write the latter rate as
\begin{eqnarray}
\widetilde{\tau}_{\Omega,X}^{-1} &=& \frac{2\pi}{\hbar} \left| M_{\text{LM,X}} \right|^2  \, \frac{\Gamma/\pi}{\left( \hbar \Omega    \right)^2 + \Gamma^2}\,, \,\,\,\,\,\,\,\,\,\,\,\,\,
\end{eqnarray}
where the photon energy $\hbar \Omega$ is measured with respect to the exciton resonance energy. The dotted line in Fig.~\ref{fig:bluex0}(c) shows the renormalized form of this expression where the broadening parameter is $\Gamma=1$~meV.  Relative to this renormalized rate, the solid lines in Fig.~\ref{fig:bluex0}(c) show the amplitude of the exchange-assisted exciton absorption from Eq.~(\ref{eq:xe2nd}) at zero temperature. The zero temperature means that the integrations in Eq.~(\ref{eq:xe2nd}) are such that $k<k_F<q$, and $|\mathbf{k}-\mathbf{q}|>k_F$. The parameters of the hBN-encapsulated WSe$_2$ monolayer are $r_0 = 4.5$~nm and $\epsilon_d = 3.8$ \cite{VanTuan_PRB18}, the exciton's Bohr radius is $a_x = 1.4$~nm \cite{Stier_PRL18},  and the effective masses are $m_h=0.36m_0$ and $m_e=0.29m_0$ \cite{Kormanyos_2DMater15}, where the translational mass of the exciton is $M=m_h+m_e=0.65m_0$. In addition, we assume parabolic energy dispersions, where the zero energy reference level is that of the exciton in the light cone at zero charge density. 

Figure~\ref{fig:Xe}(a) and (b) show the results when using the RPA screened potential and the unscreened potential, respectively (Fig.~\ref{fig:Xe}(b) is the same as Fig.~\ref{fig:bluex0}(c) in the main text). The absorption rates are about 3 times smaller when using RPA screening. Whether one uses the screened or unscreened potentials, the exchange-assisted process remains far stronger than the direct one. This dominance is shown in Fig.~\ref{fig:Xe}(c), where the exchange-assisted rate (i.e., neglecting $D({\bf q})$) is nearly three orders of magnitude larger than the direct-assisted rate (i.e., neglecting $X({\bf k,q})$). For completeness, we have also verified that the exchange-assisted absorption is far more dominant than the phonon-assisted absorption. Repeating the same procedure for the acoustic-phonon-assisted rate, we get 
\begin{eqnarray}
\tau_{\Omega,\text{ph}}^{-1} &=& \frac{2\Gamma}{\hbar} \sum_{\mathbf{q}} \left|  \frac{   M_{\text{LM,X}} \, M_{\text{X-ph}}(\mathbf{q}) }{ \varepsilon_{X, \mathbf{q}} + \hbar \omega_\mathbf{q} + i\Gamma   } \right|^2  \nonumber \\ &\times&\frac{ 1}{
\left( \hbar \Omega  -  \varepsilon_{X, \mathbf{q}} - \hbar \omega_\mathbf{q} \right)^2 + \Gamma^2 }\,\, . \label{eq:xph2nd}
\end{eqnarray}
$\hbar \omega_\mathbf{q} = \hbar v_q q$ is the phonon energy, where $v_q=3.3 \times 10^5$~cm/s is the speed of sound in the WSe$_2$ monolayer. Considering only phonon emission processes (e.g., at zero temperature), the exciton-phonon matrix element reads \cite{Yang_PRB20}
\begin{eqnarray}
M_{\text{X-ph}}(\mathbf{q}) &=&  \sqrt{ \frac{\hbar A_u}{2A v_s (2M_\text{W}+M_{\text{Se}})  }}   \nonumber \\ 
&\,&\!\!\!\!\!\! \!\!\! \!\!\! \!\!\! \!\!\! \!\!\! \!\!\! \!\!\! \!\!\! \!\!\! \!\!\! \!\!\! \!\!\! \!\!\!  \times \left( \frac{ \Xi_{e} \sqrt{q}    }{   \left(1+\left(\tfrac{1}{2}\bar{\eta} q a_x \right)^2    \right)^{3/2} }  - \frac{ \Xi_{h}  \sqrt{q}  }{  \left(1+\left(\tfrac{1}{2}\eta q a_x \right)^2    \right)^{3/2} } \right).
 \label{eq:MXp}
\end{eqnarray}
$A_u = 8.87\times 10^{-16}$~cm$^2$ is the crystal unit cell area, $M_\text{W}=3\times 10^{-22}$~g is the mass of the tungsten atom, $M_\text{Se}=1.3\times 10^{-22}$~g is the mass of the selenium atom, and the conduction- and valence-band deformation potentials are $\Xi_{e}=3.2$~eV and $\Xi_{h}=2.1$~eV, respectively \cite{Jin_PRB14}. Using Eq.~(\ref{eq:xph2nd}), the dashed-dotted line in Fig.~\ref{fig:Xe}(c) shows the relative absorption rate of the exciton assisted by emission of acoustic phonons. When the hole density is 0.5$\times$10$^{12}$~cm$^{-2}$, the exchange-assisted process is more than two orders of magnitude stronger than the acoustic-phonon-assisted absorption. 
}

\section{Exchange scattering of 2$s$ exciton states}\label{sec:exc_scat}
\begin{figure}[t!]
\centering
\includegraphics[width=8.5cm]{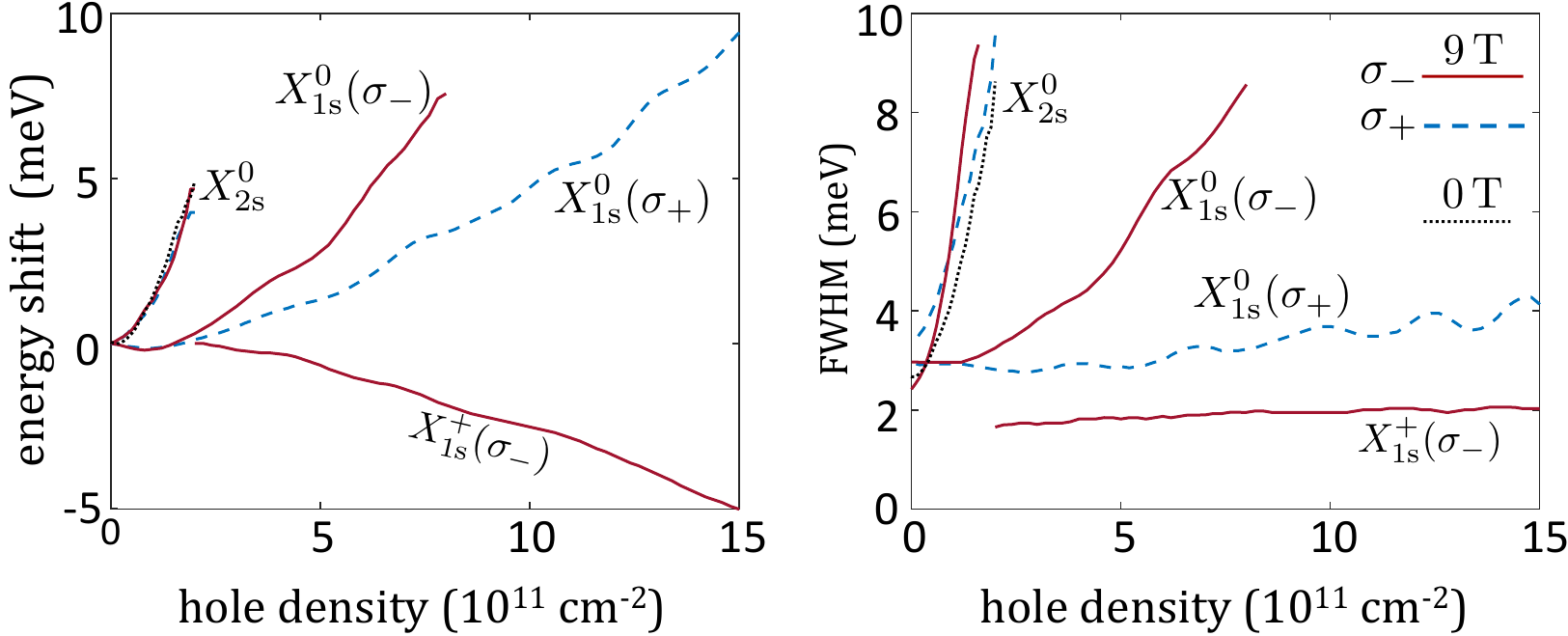}
\caption{Same as in Fig.~\ref{fig:1s}(c) and (d), but with adding the energy shift and FWHM of the $2s$ neutral exciton resonance (dotted lines are its results at zero magnetic field). The energy shift and FWHM of each resonance are shown in a range of charge densities at which the resonance is still identifiable in the reflectance spectra.}\label{fig:1s2s} 
\end{figure}

Figure~\ref{fig:1s2s} shows the energy shift and FWHM of exciton resonances as a function of hole density in a hBN-encapsulated WSe$_2$ monolayer. The results are the same as in Fig.~\ref{fig:1s}, where here we add the corresponding results of the first excited state of the neutral exciton ($2s$). This resonance emerges $\sim$130~meV above that of the ground state exciton ($1s$) \cite{Stier_PRL18,VanTuan_PRB18}.  We notice that the FWHM and energy shifts of the $2s$ excited state are stronger than those of the $1s$ ground state, which we  attribute to the larger exciton size in the $2s$ state \cite{Stier_PRL18}. Simply put, excitons with larger radii are more sensitive to charge density changes in their vicinity \cite{VanTuan_PRB23}.%

More interestingly, Fig.~\ref{fig:1s2s} shows that optimality only affects the blueshift and broadening of the exciton resonance in its ground state. In more detail, the discussion in the main text explained why the resonance of an optimal exciton shifts and broadens less in energy, evidenced by comparing the behavior of $X^0_{\text{1s}}(\sigma_\pm)$. Yet, the corresponding lines that are labeled by $X^0_{\text{2s}}$ show no regard to whether the exciton is distinguishable (solid line; $\sigma_-$ at 9$\,$T), optimal (dashed line; $\sigma_+$ at 9$\,$T), or neither (dotted line; 0$\,$T). To settle this apparent contradiction we should consider that excited state excitons are subjected to ultrafast energy relaxation through exchange scattering. As shown in Ref.~\cite{Yang_PRB22}, an exciton created in the $2s$ state is turned to a hot exciton in the $1s$ state as fast as 0.1~ps upon exchange scattering with a resident carrier when the charge density is $\sim10^{11}$~cm$^{-2}$. Such timescales reinforce the larger broadening of excited-state exciton resonances in Fig.~\ref{fig:1s2s}. Whether the resident carrier with which the exciton interacts is distinguishable or indistinguishable determines if the exchange interaction is attractive or repulsive  \cite{VanTuan_PRB25}. Yet, the energy relaxation rate is similar in both cases because the square amplitude of the exchange-scattering matrix element is indifferent to this change in sign \cite{Yang_PRB22}. As a result, the rate at which an exciton can relax from an excited state to the ground state depends on charge density and not on  distinguishability of the resident carriers \cite{trions_2s}.

{\color{black}\section{Electron-hole exchange interaction}\label{sec:eh_exc}

The analysis we have presented in this work keeps its generality in the presence of the e-h exchange interaction. The latter is not the exciton-electron exchange interaction discussed before, but a repulsive interaction between an electron in the conduction band and a missing electron (hole) in the valence band with the same spin \cite{Pikus_Bir_71,Maialle_PRB93}. The amplitude of this interaction is commensurate with the spatial overlap between the electron and hole, and the relatively tight overlap in TMD monolayers is such that the long-range part of the e-h exchange interaction is responsible for the fast valley depolarization of bright excitons in these materials \cite{Yu_NatComm14,Yu_PRB14,Glazov_PSSB15,Yang_PRB20}.

\vspace{2mm}

The short-range part of the e-h exchange interaction lifts the energy degeneracy between various excitonic complexes \cite{Qiu_PRL15,Steinhoff_NatPhys18}. In tungsten-based monolayers, for example, the repulsive exchange interaction between an hole and electron in opposite valleys increases the energy of the momentum-indirect exciton with respect to the energy of the dark trion \cite{Li_PRB22}, or of the triplet negative trion with respect to the singlet negative trion \cite{Hichri_PRB20}. Importantly, the fine-structure splittings are largely insensitive to changes in charge density because of the short range nature of the interaction that introduces these splittings. For example, the energy splitting between $X^-_T$ and $X^-_S$ in a WSe$_2$ monolayer remains the same when the charge density increases, as shown in Fig.~\ref{fig:wse2case}(b). Their broadening  is also similar, as shown in Fig.~\ref{fig:wse2case}(c), providing additional support that the e-h exchange interaction is indifferent to changes in charge density.

}


\end{document}